# Interest Rates After The Credit Crunch: Multiple-Curve Vanilla Derivatives and SABR

*Version: Friday, 28 March 2012*

## Marco Bianchetti

Market Risk Management, Intesa Sanpaolo Bank, Piazza Paolo Ferrari, 10, 20121 Milan, Italy,
marco.bianchetti[AT]intesasanpaolo.com

## Mattia Carlicchi

Market Risk Management, Intesa Sanpaolo Bank, Piazza Paolo Ferrari, 10, 20121 Milan, Italy,
mattia.carlicchi[AT]intesasanpaolo.com

## Abstract

We present a quantitative study of the markets and models evolution across the credit crunch crisis. In particular, we focus on the fixed income market and we analyze the most relevant empirical evidences regarding the divergences between Libor and OIS rates, the explosion of Basis Swaps spreads, and the diffusion of collateral agreements and CSA-discounting, in terms of credit and liquidity effects.

We also review the new modern pricing approach prevailing among practitioners, based on multiple yield curves reflecting the different credit and liquidity risk of Libor rates with different tenors and the overnight discounting of cash flows originated by derivative transactions under collateral with daily margination. We report the classical and modern no-arbitrage pricing formulas for plain vanilla interest rate derivatives, and the multiple-curve generalization of the market standard SABR model with stochastic volatility.

We then report the results of an empirical analysis on recent market data comparing pre- and post-credit crunch pricing methodologies and showing the transition of the market practice from the classical to the modern framework. In particular, we prove that the market of Interest Rate Swaps has abandoned since March 2010 the classical Single-Curve pricing approach, typical of the pre-credit crunch interest rate world, and has adopted the modern Multiple-Curve CSA approach, thus incorporating credit and liquidity effects into market prices. The same analysis is applied to European Caps/Floors, finding that the full transition to the modern Multiple-Curve CSA approach has retarded up to August 2010. Finally, we show the robustness of the SABR model to calibrate the market volatility smile coherently with the new market evidences.

## Acknowledgments

The authors gratefully acknowledge fruitful interactions with A. Battauz, A. Castagna, C. C. Duminuco, F. Mercurio, M. Morini, M. Trapletti and colleagues at Market Risk Management and Fixed Income trading desks.

# 1. Introduction

The financial crisis begun in the second half of 2007 has triggered, among many consequences, a deep evolution phase of the classical framework adopted for trading derivatives. In particular, credit and liquidity issues were found to have macroscopical impacts on the prices of financial instruments, both plain vanillas and exotics. Today, terminated or not the crisis, the market has learnt the lesson and persistently shows such effects. These are clearly visible in the market quotes of plain vanilla interest rate derivatives, such as Deposits, Forward Rate Agreements (FRA), Swaps (IRS) and options (Caps, Floors and Swaptions). Since August 2007 the primary interest rates of the interbank market, e.g. Libor, Euribor, Eonia, and Federal Funds rate[1], display large basis spreads that have raised up to 200 basis points. Similar divergences are also found between FRA rates and the forward rates implied by two consecutive Deposits, and similarly, among swap rates with different floating leg tenors. Recently, the market has also included the effect of collateral agreements widely diffused among derivatives counterparties in the interbank market.

After the market evolution the standard no-arbitrage framework adopted to price derivatives, developed over forty years following the Copernican Revolution of Black and Scholes (1973) and Merton (1973), became obsolete. Familiar relations described on standard textbooks (see e.g. Brigo and Mercurio (2006), Hull (20010)), such as the basic definition of forward interest rates, or the swap pricing formula, had to be abandoned. Also the fundamental idea of the construction of a single risk free yield curve, reflecting at the same time the present cost of funding of future cash flows and the level of forward rates, has been ruled out. The financial community has thus been forced to start the development of a new theoretical framework, including a larger set of relevant risk factors, and to review from scratch the no-arbitrage models used on the market for derivatives' pricing and risk analysis. We refer to such old and new frameworks as "classical" and "modern", respectively, to remark the shift of paradigm induced by the crisis.

The paper is organised as follows. In section 2 we describe the market evolution, focusing on interest rates, and we discuss in detail the empirical evidences regarding basis spreads and collateral effects cited above. In section 3 we focus on the methodological evolution from the classical to the modern pricing framework, describing the foundations of the new multiple yield curves framework adopted by market practitioners in response to the crisis. In section 4 we report the results of an empirical analysis on recent market data comparing three different pre- and post-credit crunch pricing methodologies, showing the transition of the market practice from the classical to the modern pricing framework. We also report a study of the SABR stochastic volatility model – the market standard for pricing and hedging plain vanilla interest rate options – showing its robustness under generalisation to the modern framework and to calibrate the market volatility surfaces across the crisis. Conclusions and directions of future works are collected in section 5.

The topics discussed here are at the heart of the present derivatives market, with many consequences in trading, financial control, risk management and IT, and are attracting a growing attention in the financial literature. To our knowledge, they have been approached by Kijima et al. (2008), Chibane and Sheldon (2009), Ametrano and Bianchetti (2009), Ametrano (2011), Fujii et al. (2009a, 2010a, 2011) in terms of multiple-curves; by Henrard (2007, 2009) and Fries (2010) using a first-principles approach; by Bianchetti (2010) using a foreign currency approach; by Fujii et al. (2009b), Mercurio (2009, 2010a, 2010b) and Amin (2010) within the Libor Market Model; by Pallavicini and Tarenghi (2010) and Moreni and Pallavicini (2010) within the HJM model; by Kenyon (2010) using a short rate model; by Morini (2009) in terms of counterparty risk; by Burghard and Kjaer (2010), Piterbarg (2010a, 2010b), Fujii et al. (2010b), Morini and Prampolini (2010) in terms of cost of funding. See also the Risk Magazine reports of Madigan (2008), Wood (2009a, 2009b) and Whittall (2010a, 2010b, 2010c).

# 2. Market Evolution

In this section we discuss the most important market data showing the main consequences of the credit crunch crisis started in August 2007. We will focus, in particular, on Euro interest rates, since they show rather peculiar and persistent effects that have strong impacts on pricing methodologies. The same results hold for other currencies, USDLibor and Federal Funds rates in particular (see. e.g Mercurio (2009, 2010b)).

---

[1] Libor, sponsored by the British Banking Association (BBA), is quoted in all the major currencies and is the reference rate for international Over-The-Counter (OTC) transactions (see www.bbalibor.com). Euribor and Eonia, sponsored by the European Banking Federation (EBF), are the reference rates for OTC transactions in the Euro market (see www.euribor.org). The Federal Funds rate is a primary rate of the USD market and is set by the Federal Open Market Committee (FOMC) accordingly to the monetary policy decisions of the Federal Reserve (FED).

## 2.1. Euribor – OIS Basis

Figure 1 reports the historical series of the Euribor Deposit 6-month (6M) rate versus the Eonia Overnight Indexed Swap[2] (OIS) 6-month (6M) rate over the time interval Jan. 06 – Dec. 10. Before August 2007 the two rates display strictly overlapping trends differing of no more than 6 bps. In August 2007 we observe a sudden increase of the Euribor rate and a simultaneous decrease of the OIS rate that lead to the explosion of the corresponding basis spread, touching the peak of 222 bps in October 2008, when Lehman Brothers filed for bankruptcy protection. Successively the basis has sensibly reduced and stabilized between 40 bps and 60 bps. Notice that the pre-crisis level has never been recovered. The same effect is observed for other similar couples, e.g. Euribor 3M vs OIS 3M.

The reason of the abrupt divergence between the Euribor and OIS rates can be explained by considering both the monetary policy decisions adopted by international authorities in response to the financial turmoil, and the impact of the credit crunch on the credit and liquidity risk perception of the market, coupled with the different financial meaning and dynamics of these rates.

- The Euribor rate is the reference rate for over-the-counter (OTC) transactions in the Euro area. It is defined as "the rate at which Euro interbank Deposits are being offered within the EMU zone by one prime bank to another at 11:00 a.m. Brussels time". The rate fixings for a strip of 15 maturities, ranging from one day to one year, are constructed as the trimmed average of the rates submitted (excluding the highest and lowest 15% tails) by a panel of banks. The Contribution Panel is composed, as of September 2010, by 42 banks, selected among the EU banks with the highest volume of business in the Euro zone money markets, plus some large international bank from non-EU countries with important euro zone operations. Thus, Euribor rates reflect the average cost of funding of banks in the interbank market at each given maturity. During the crisis the solvency and solidity of the whole financial sector was brought into question and the credit and liquidity risk and premia associated to interbank counterparties sharply increased. The Euribor rates immediately reflected these dynamics and raise to their highest values over more than 10 years. As seen in Figure 1, the Euribor 6M rate suddenly increased on August 2007 and reached 5.49% on 10$^{th}$ October 2008.

- The Eonia rate is the reference rate for overnight OTC transactions in the Euro area. It is constructed as the average rate of the overnight transactions (one day maturity deposits) executed during a given business day by a panel of banks on the interbank money market, weighted with the corresponding transaction volumes. The Eonia Contribution Panel coincides with the Euribor Contribution Panel. Thus Eonia rate includes information on the short term (overnight) liquidity expectations of banks in the Euro money market. It is also used by the European Central Bank (ECB) as a method of effecting and observing the transmission of its monetary policy actions. During the crisis the central banks were mainly concerned about restabilising the level of liquidity in the market, thus they reduced the level of the official rates: the "Deposit Facility rate" and the "Marginal Lending Facility rate". This is clear from Figure 2, showing that, over the period Jan. 06 – Dec. 10, Eonia is always higher than the Deposit Facility rate and lower than the Marginal Lending Facility rate, defining the so-called "Rates Corridor". Furthermore, the daily tenor of the Eonia rate makes negligible the credit and liquidity risks reflected on it: for this reason the OIS rates are considered the best proxies available in the market for the risk-free rate.

---

[2] The Overnight Index Swap (OIS) is a swap with a fixed leg versus a floating leg indexed to the overnight rate. The Euro market quotes a standard OIS strip indexed to Eonia rate (daily compounded) up to 30 years maturity.

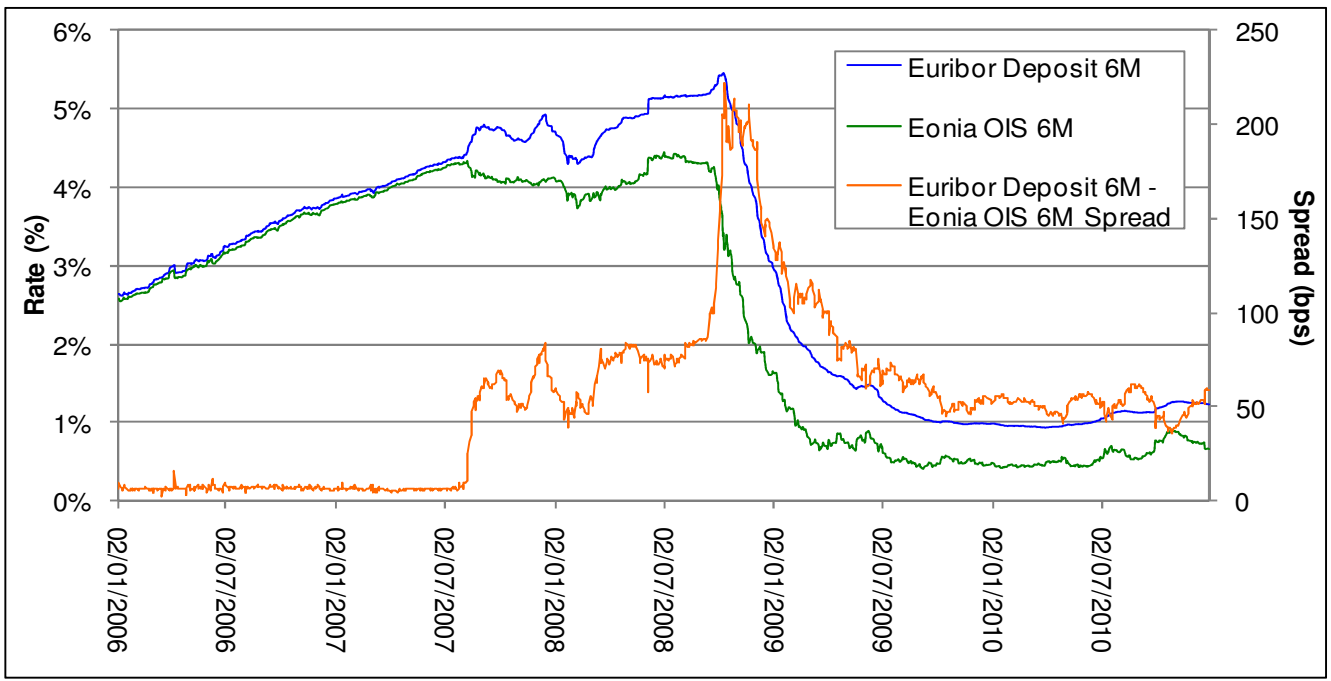

**Figure 1:** historical series of Euribor Deposit 6M rate versus Eonia OIS 6M rate. The corresponding spread is shown on the right axis (Jan. 06 – Dec. 10 window, source: Bloomberg).

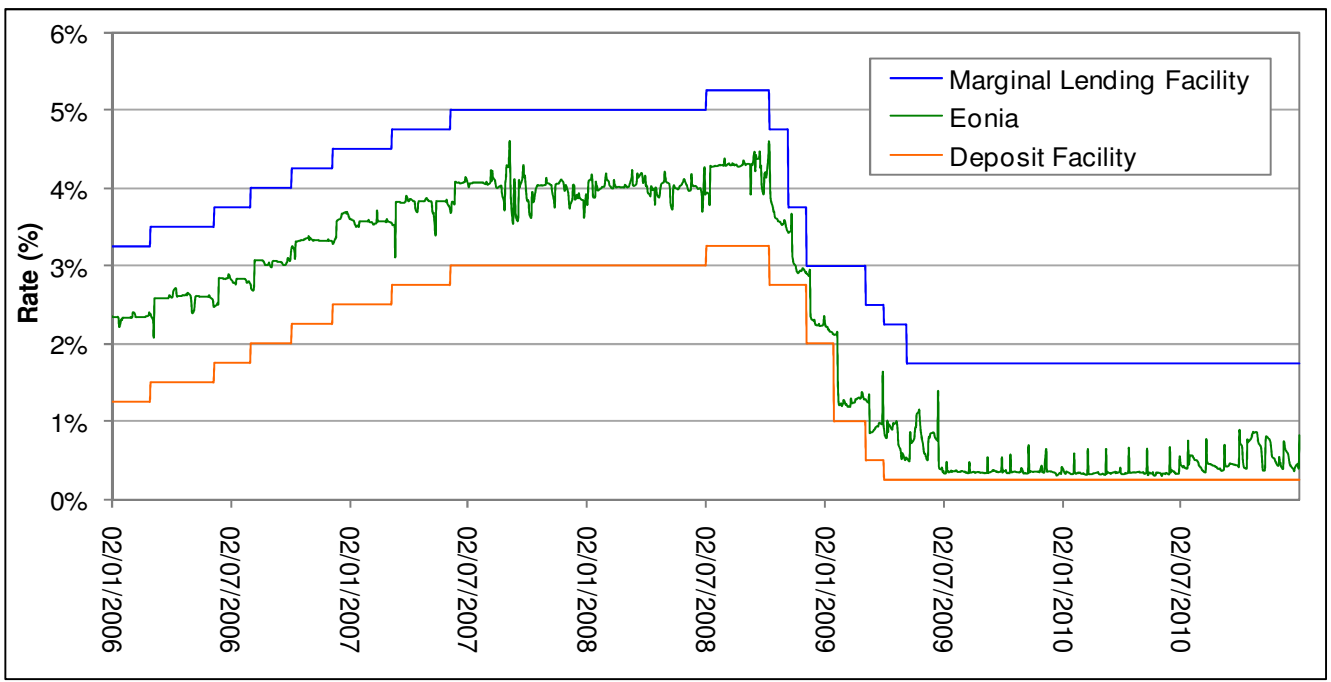

**Figure 2:** historical series of the Deposit Lending Facility rate, of the Marginal Lending Facility rate and of the Eonia rate (Jan. 06 – Dec. 10 window, sources: European Central Bank – Press Releases and Bloomberg).

Thus the Euribor-OIS basis explosion of August 2007 plotted in Figure 1 is essentially a consequence of the different credit and liquidity risk reflected by Euribor and Eonia rates. We stress that such divergence is not a consequence of the counterparty risk carried by the financial contracts, Deposits and OISs, exchanged in the interbank market by risky counterparties, but depends on the different fixing levels of the underlying Euribor and Eonia rates.

The different influence of credit risk on Libor and overnight rates can be also appreciated in Figure 3, where we compare the historical series for the Euribor-OIS spread of Figure 1 with those of Credit Default Swaps (CDS) spreads for some main banks in the Euribor Contribution Panel. We observe that the Euribor-OIS basis explosion of August 2007 exactly matches the CDS explosion, corresponding to the generalized increase of the default risk seen in the interbank market.

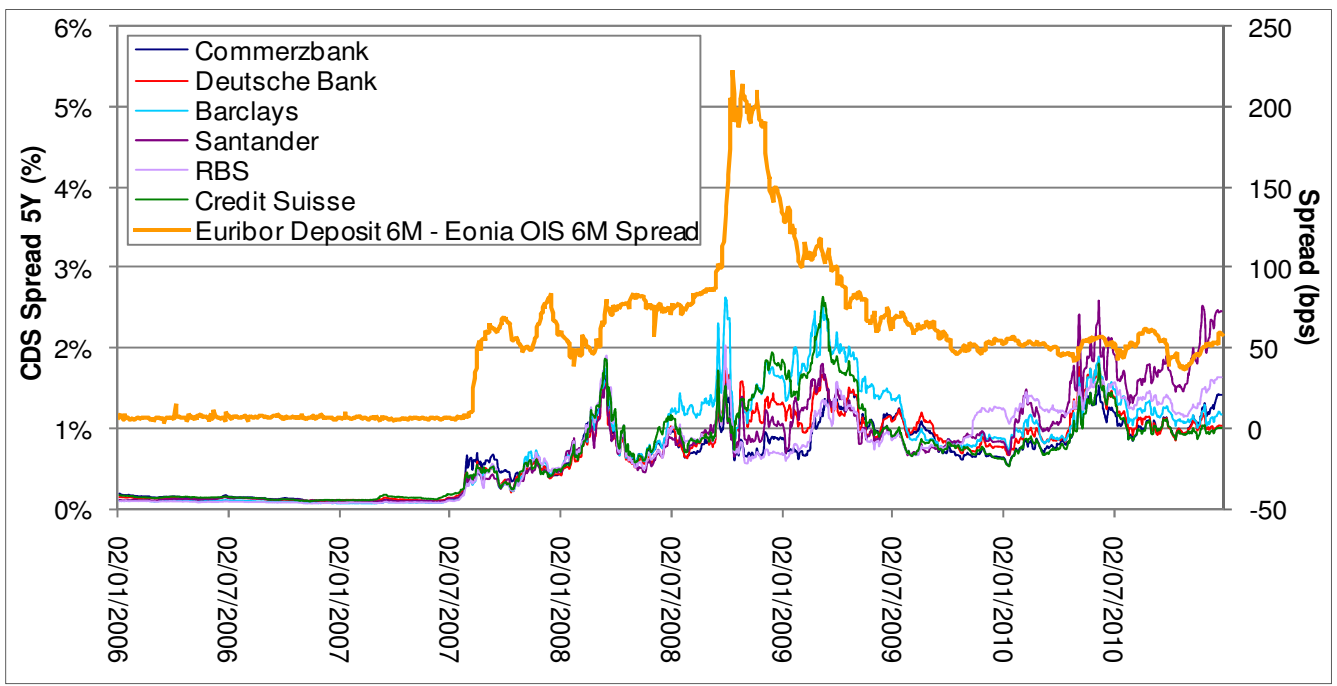

**Figure 3:** left y-axis: CDS Spread 5Y for some European banks belonging to the Euribor panel. Right y-axis: spread between the Euribor Deposit 6M – Eonia OIS 6M from Figure 1 (Jan. 06 – Dec. 10 window, source: Bloomberg).

The liquidity risk component in Euribor and Eonia interbank rates is distinct but strongly correlated to the credit risk component. According to Acerbi and Scandolo (2007), liquidity risk may appear in at least three circumstances:

1. lack of liquidity to cover short term debt obligations (funding liquidity risk),
2. difficulty to liquidate assets on the market due excessive bid-offer spreads (market liquidity risk),
3. difficulty to borrow funds on the market due to excessive funding cost (systemic liquidity risk).

Following Morini (2009), these three elements are, in principle, not a problem until they do not appear together, because a bank with, for instance, problem 1 and 2 (or 3) will be able to finance itself by borrowing funds (or liquidating assets) on the market. During the crisis these three scenarios manifested themselves jointly at the same time, thus generating a systemic lack of liquidity (see e.g. Michaud and Upper (2008)).

Clearly, it is difficult to disentangle liquidity and credit risk components in the Euribor and Eonia rates, because, in particular, they do not refer to the default risk of one counterparty in a single derivative deal but to a money market with bilateral credit risk (see the discussion in Morini (2009) and references therein).

Finally, we stress that, as seen in Figure 1, the Libor-OIS basis is still persistent today at a non-negligible level, despite the lower rate and higher liquidity regime reached after the most acute phase of the crisis and the strong interventions of central banks and governments. Clearly the market has learnt the lesson of the crisis and has not forgotten that these interest rates are driven by different credit and liquidity dynamics. From an historical point of view, we can compare this effect to the appearance of the volatility smile on the option markets after the 1987 crash (see e.g. Derman and Kani (1994)). It is still there.

## 2.2. FRA Rates versus Forward Rates

The considerations above, referred to spot Euribor and Eonia rates underlying Deposit and OIS contracts, apply to forward rates as well. In Figure 4 we report the historical series of quoted Euribor Forward Rate Agreement (FRA) 3x6 rates versus the forward rates implied by the corresponding Eonia OIS 3M and 6M rates. The FRA 3x6 rate is the equilibrium (fair) rate of a FRA contract starting at spot date (today + 2 working days in the Euro market), maturing in 6 months, with a floating leg indexed to the forward interest rate between 3 and 6 months, versus a fixed interest rate leg. The paths of market FRA rates and of the corresponding forward rates implied in two consecutive Eonia OIS Deposits observed in Figure 4 are similar to those observed in Figure 1 for the Euribor Deposit and Eonia OIS respectively. In particular, a sudden divergence between the quoted FRA rates and the implied forward rates arose in August 2007, regardless the maturity, and reached its peak in October 2008 with the Lehman crash.

Mercurio (2009) has proven that the effects above may be explained within a simple credit model with a default-free zero coupon bond and a risky zero coupon bond emitted by a defaultable counterparty with recovery rate *R*. The associated risk free and risky Libor rates are the underlyings of the corresponding risk free and risky FRAs.

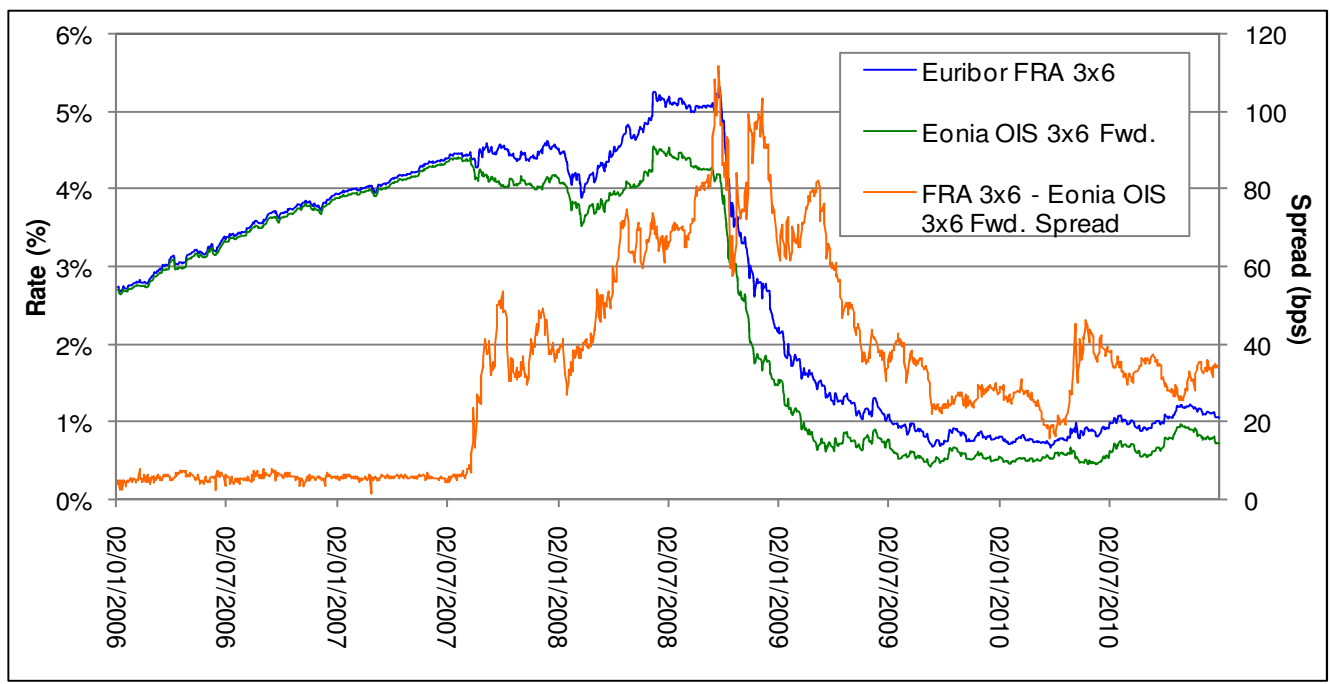

**Figure 4:** FRA 3x6 market quote versus 3 months forward rate implied in two consecutive 3M and 6M Eonia OIS rates. The corresponding spread is shown on the right y-axis (Jan. 06 – Dec. 10 window, source: Bloomberg).

## 2.3. Basis Swaps

A third evidence of the regime change after the credit crunch is the explosion of the Basis Swaps spreads. In Figure 5 we report three historical series of quoted Basis Swap equilibrium spread, Euribor 3M vs Euribor 6M, Euribor 6M vs Euribor 12M, Euribor 3M vs Eonia, all at 5 years swap maturity. Basis Swaps are quoted on the Euro interbank market in terms of the difference between the fixed equilibrium swap rates of two swaps. For instance, the quoted Euribor 3M vs Euribor 6M Basis Swap rate is the difference between the fixed rates of a first standard swap with an Euribor 3M floating leg (quarterly frequency) vs a fixed leg (annual frequency), and of a second swap with an Euribor 6M floating leg (semi-annual frequency) vs a fixed leg (annual frequency). The frequency of the floating legs is the "tenor" of the corresponding Euribor rates. The Eonia rate has the shortest tenor (1 day). As we can see in Figure 5, the basis swap spreads were negligible (or even not quoted) before the crisis. They suddenly diverged in August 2007 and peaked in October 2008 with the Lehman crash.

The Basis Swap involves a sequence of spot and forward rates carrying the credit and liquidity risk discussed in sections 2.1 and 2.2 above. Hence, the basis spread explosion can be interpreted, in principle, in terms of the different credit and liquidity risk carried by the underlying Libor rates with different tenor. From the market evidences reported in Figure 5 we understand that, after the crisis, market players have a preference for receiving floating payments with higher frequency (e.g. 3M) indexed to lower tenor Euribor rates (e.g. Euribor 3M), with respect to floating payments with lower frequency (e.g. 6M) indexed to higher tenor Euribor rates (e.g. Euribor 6M), and are keen to pay a premium for the difference. Hence in a Basis Swap (e.g. 3M vs 6M) the floating leg indexed to the higher rate tenor (6M) must include a risk premium higher than that included in the floating leg indexed to the shorter rate tenor (3M, both with the same maturity). Thus a positive spread emerges between the two corresponding equilibrium rates (or, in other words, a positive spread must be added to the 3M floating leg to equate the value of the 6M floating leg).

According to Morini (2009), a basis swap between two interbank counterparties under collateral agreement can be described as the difference between two investment strategies. Fixing, for instance, a Basis Swap Euribor 3M vs Euribor 6M with 6M maturity, scheduled on 3 dates $T_0$, $T_1=T_0+3M$, $T_2=T_0+6M$, we have the following two strategies:

1. 6M floating leg: at $T_0$ choose a counterparty $C_1$ with an high credit standing (that is, belonging to the Euribor Contribution Panel) with collateral agreement in place, and lend the notional for 6 months at the Euribor 6M rate prevailing at $T_0$ (Euribor 6M flat because $C_1$ is an Euribor counterparty). At maturity $T_2$ recover notional plus interest from $C_1$. Notice that if counterparty $C_1$ defaults within 6 months we gain full recovery thanks to the collateral agreement.

2. 3M+3M floating leg: at $T_0$ choose a counterparty $C_1$ with an high credit standing (belonging to the Euribor Contribution Panel) with collateral agreement in place, and lend the notional for 3 months at the Euribor 3M rate (flat) prevailing at $T_0$. At $T_1$ recover notional plus interest and check the credit standing of $C_1$: if $C_1$ has maintained its credit standing (it still belongs to the Euribor Contribution Panel), then lend the money again to $C_1$ for 3 months at the Euribor 3M rate (flat) prevailing at $T_1$, otherwise choose another counterparty $C_2$ belonging to the Euribor Panel with collateral agreement in place, and lend the money to $C_2$ at the same interest rate. At maturity $T_2$ recover notional plus interest from $C_1$ or $C_2$. Again, if counterparties $C_1$ or $C_2$ defaults within 6 months we gain full recovery thanks to the collateral agreements.

Clearly, the 3M+3M leg implicitly embeds a bias towards the group of banks with the best credit standing, typically those belonging to the Euribor Contribution Panel. Hence the counterparty risk carried by the 3M+3M leg must be lower than that carried by the 6M leg. In other words, the expectation of the survival probability of the borrower of the 3M leg in the second 3M-6M period is higher than the survival probability of the borrower of the 6M leg in the same period. This lower risk is embedded into lower Euribor 3M rates with respect to Euribor 6M rates. But with collateralization the two legs have both null counterparty risk. Thus a positive spread must be added to the 3M+3M leg to reach equilibrium. The same discussion can be repeated, mutatis mutandis, in terms of liquidity risk.

We stress that the credit and liquidity risk involved here are those carried by the risky Libor rates underlying the Basis Swap, reflecting the average default and liquidity risk of the interbank money market (of the Libor panel banks), not those associated to the specific counterparties involved in the financial contract. We stress also that such effects were already present before the credit crunch, as discussed e.g. in Tuckman and Porfirio (2004), and well known to market players, but not effective due to negligible basis spreads.

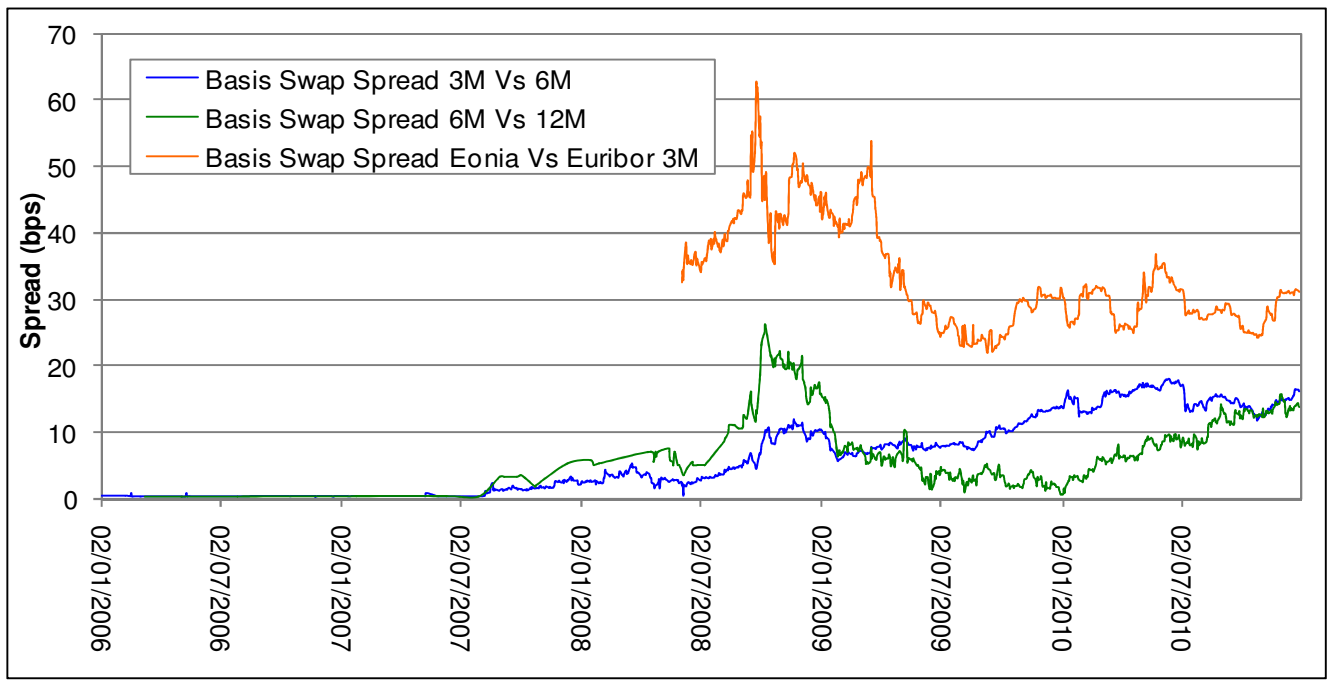

**Figure 5:** Basis Swap spreads: Euribor 3M Vs Euribor 6M, Euribor 6M Vs Euribor 12M and Eonia Vs Euribor 3M (Jan. 06 – Dec. 10 window, source: Bloomberg). Notice that the daily market quotations for some basis swap were not even available before the crisis.

## 2.4. Collateralization and OIS-Discounting

Another effect of the credit crunch has been the great diffusion of collateral agreements to reduce the counterparty risk of OTC derivatives positions. Nowadays most of the counterparties on the interbank market have mutual collateral agreements in place. In 2010, more than 70% of all OTC derivatives transactions were collateralized (ISDA (2010)).

Typical financial transactions generate streams of future cash flows, whose total net present value (NPV = algebraic sum of all discounted expected cash flows) implies a credit exposure between the two counterparties. If, for counterparty A, NPV(A)>0, then counterparty A expects to receive, on average, future cash flows from counterparty B (in other words, A has a credit with B). On the other side, if counterparty B has NPV(B)<0, then it expects to pay, on average, future cash flows to counterparty A (in other words, B has a debt with A). The reverse holds if NPV(A)<0 and NPV(B)>0. Such credit exposure can be mitigated through a guarantee, called "collateral agreement", or " Credit Support Annex" (CSA), following the International Swaps and Derivatives Association (ISDA) standards widely used to regulate OTC transactions. The main feature of the CSA is a margination mechanism similar to those adopted by central clearing houses for standard instruments exchange (e.g. Futures). In a nutshell, at every margination date the two counterparties check the value of the portfolio of mutual OTC transactions and regulate the margin, adding to or subtracting from the collateral account the corresponding mark to market variation with respect to the preceding margination date. The margination can be regulated with cash or with (primary) assets of corresponding value. In any case the collateral account holds, at each date, the total NPV of the portfolio, which is positive for the creditor counterparty and negative for the debtor counterparty. The collateral amount is available to the creditor. On the other side, the debtor receives an interest on the collateral amount, called "collateral rate". Hence, we can see the collateral mechanism as a funding mechanism, transferring liquidity from the debtor to the creditor. The main differences with traditional funding through Deposit contracts are that, using derivatives, we have longer maturities and stochastic lending/borrowing side and amount. We can also look at CSA as an hedging mechanism, where the collateral amount hedges the creditor against the event of default of the debtor. The most diffused CSA provides a daily margination mechanism and an overnight collateral rate (ISDA (2010)). Actual CSAs provide many other detailed features that are out of the scope of the present discussion.

Thus, a first important consequence of the diffusion of collateral agreements among interbank counterparties is that we can consider the derivatives' prices quoted on the interbank market as counterparty risk free OTC transactions. A second important consequence is that, by no-arbitrage, the CSA margination rate and the discounting rate of future cash flows must match. Hence the name of "CSA discounting". In particular, the most diffused overnight CSA implies overnight-based discounting and the construction of a discounting yield curve that must reflect, for each maturity, the funding level in an overnight collateralized interbank market. Thus Overnight Indexed Swaps (OIS) are the natural instruments for the discounting curve construction. Hence the alternative name of "OIS discounting" or "OIS (yield) curve". Such discounting curve is also the best available proxy of a risk free yield curve.

In case of absence of CSA, using the same no-arbitrage principle between the funding and the discounting rate, we conclude that a bank should discount future cash flows (positive or negative) using its own "traditional" cost of funding term structure. This implies important (and rather involved) consequences, such that, according to Morini and Prampolini (2009), each counterparty assigns a different present value to the same future cash flow, breaking the fair value symmetry; that a worsening of the its credit standing allows the Bank to sell derivatives (options in particular) at more competitive prices (the lower the rate, the higher the discount, the lower the price); the problem of double counting the Debt Value Adjustment (DVA) to the fair value.

Presently, the market is in the middle of a transition phase from the classical Libor-based discounting methodology to the modern CSA-based discounting methodology. OTC transactions executed on the interbank market normally use CSA discounting. In particular, plain vanilla interest rate derivatives, such as FRA, Swaps, Basis Swaps, Caps/Floor/Swaptions are quoted by main brokers using CSA discounting (ICAP (2010)). On the other side, presently just a few banks have declared full adoption of CSA discounting also for balance sheet revaluation and collateral margination (see e.g. Bianchetti (2011)).

Finally we stress that also before the crisis the old-style standard Libor curve was representative of the average funding level on the interbank market (see e.g. Hull (2010)). Such curve, even if considered a good proxy for a risk free curve, thanks to the perceived low counterparty risk of primary banks (belonging to the Libor Contribution panel), was not strictly risk free because of the absence of collateralization.

# 3. Modelling Evolution

According to Bianchetti and Morini (2010), the market "frictions" discussed in sec. 2 have induced a sort of "segmentation" of the interest rate market into sub-areas, mainly corresponding to instruments with 1M, 3M, 6M, 12M underlying rate tenors. These are characterized, in principle, by different internal dynamics, liquidity and credit risk premia, reflecting the different views and interests of the market players. In response to the crisis, the classical pricing framework, based on a single yield curve used to calculate forward rates and discount factors, has been abandoned, and a new modern pricing approach has prevailed among practitioners. The new methodology takes into account the market segmentation as an empirical evidence and incorporates the new interest rate dynamics into a multiple curve framework as follows.

- **Discounting curves**: these are the yield curves used to discount futures cash flows. As discussed in section 2.4, the curve must be constructed and selected such that to reflect the cost of funding of the bank in connection with the actual nature of the specific contract that generates the cash flows. In particular:

    o an OIS-based curve is used to discount cash flow generated by a contract under CSA with daily margination and overnight collateral rate;

    o a funding curve is used in case of contracts without CSA;

    o in case of non-standard CSA (e.g. different margination frequency, rate, threshold, etc.), appropriate curves should be, in principle, selected, but we will not discuss this topic here since it applies to a minority of deals and it would be out of the scope of the present paper.

    We stress that the funding curve for no-CSA contracts is specific to each counterparty, that will have its specific funding curve. This modern discounting methodology is called CSA-discounting.

- **Forwarding curves**: these are the yield curves used to compute forward rates. As discussed in section 2.3, the curve must be constructed and selected according to the tenor and typology of the rate underlying the actual contract to be priced. For instance, a Swap floating leg indexed to Euribor 6M requires the an Eurbor 6M forwarding curve constructed from quoted instruments with Euribor 6M underlying rate.

Following Bianchetti (2010), we report in Table 1 the comparison between the classical and the modern frameworks, called Single-Curve and Multiple-Curve approach, respectively.

The adoption of the Multiple-Curve approach has led to the revision of no-arbitrage pricing formulas. According to Mercurio (2009, 2010a, 2010b), we compare in Table 2 the classical and modern pricing formulas for plain vanilla interest rate derivatives.

We stress that the fundamental quantity of the modern pricing framework is the FRA rate $\tilde{F}$. Indeed, following Mercurio (2009, 2010a, 2010b), the correct probability measure to be used into expectations is that associated to the discounting curve $\mathcal{C}_d$, under which the forward rate is no longer a martingale. Instead, the FRA rate, by definition, is a martingale under such measure.

|  | **Classical Methodology (Single-Curve)** | **Modern Methodology (Multiple-Curve)** |
|---|---|---|
| **1. Yield curves construction** | Select a single finite set of the most convenient (e.g. liquid) vanilla interest rate market instruments and build a single yield curve $\mathcal{C}$ using the preferred bootstrapping procedure. For instance, a common choice in the European market is a combination of short-term EUR deposits, medium-term Futures/FRAs on Euribor 3M and medium-long-term swaps on Euribor 6M. | Build one discounting curve $\mathcal{C}_d$ using the preferred selection of vanilla interest rate market instruments and bootstrapping procedure.<br><br>Build multiple distinct forwarding curves $\mathcal{C}_x$ using the preferred selections of distinct sets of vanilla interest rate market instruments, each homogeneous in the underlying rate tenor (typically $x = 1M, 3M, 6M, 12M$) and bootstrapping procedures. For example, for the construction of the forwarding curve $\mathcal{C}_{6M}$ only market instruments with 6-month tenor are considered. |
| **2. Computation of expected cash flows** | For each interest rate coupon compute the relevant forward rates using the given yield curve $\mathcal{C}$ and applying the standard formula, with $t \leq T_{k-1} \leq T_k$:<br>$$F_k(t) := F(t; T_{k-1}, T_k) = \frac{1}{\tau_k}\left[\frac{P(t, T_{k-1})}{P(t, T_k)} - 1\right],$$<br>where $\tau_k$ is the year fraction related to the time interval $[T_{k-1}, T_k]$.<br>Compute cash flows as expectations at time $t$ of the corresponding coupon payoffs with respect to the $T_k$-forward measure $Q^{T_k}$, associated to the numeraire $P(t, T_k)$ from the same yield curve $\mathcal{C}$:<br>$$c_k(t, T_k, \pi_k) = E^{Q^{T_k}}[\pi_k | F_t].$$ | For each interest rate coupon compute the relevant FRA rate $\tilde{F}_{x,k}(t)$ with tenor $x$ using the corresponding forwarding curve $\mathcal{C}_x$ and applying the following formula, with $t \leq T_{k-1} \leq T_k$:<br>$$\tilde{F}_{x,k}(t) := \tilde{F}_x(t; T_{k-1}, T_k) = \frac{1}{\tau_{x,k}}\left[\frac{P_x(t, T_{k-1})}{P_x(t, T_k)} - 1\right],$$<br>where $\tau_{x,k}$ is the year fraction associated to the time interval $[T_{k-1}, T_k]$.<br>Compute cash flows as expectations at time $t$ of the corresponding coupon payoffs with respect to the discounting $T_k$-forward measure $Q_d^{T_k}$, associated to the numeraire $P_d(t, T_k)$ from the discounting curve $\mathcal{C}_d$:<br>$$c_k(t, T_k, \pi_k) = E^{Q_d^{T_k}}[\pi_k | F_t].$$ |
| **3. Computation of discount factors** | Compute the relevant discount factors $P(t, T_k)$ from the unique yield curve $\mathcal{C}$ defined in step 1. | Compute the relevant discount factors $P_d(t, T_k)$ from the discounting curve $\mathcal{C}_d$ of step 1. |
| **4. Computation of the derivative's price** | Compute the derivative's price at time $t$ as the sum of the discounted expected future cash flows:<br>$$V(t) = \sum_{k=1}^{m} P(t, T_k) c_k(t, T_k, \pi_k)$$<br>$$= \sum_{k=1}^{m} P(t, T_k) \mathbb{E}^{Q^{T_k}}[\pi_k | \mathcal{F}_t]$$ | Compute the derivative's price at time $t$ as the sum of the discounted expected future cash flows:<br>$$V(t) = \sum_{k=1}^{m} P_d(t, T_k) c_k(t, T_k, \pi_k)$$<br>$$= \sum_{k=1}^{m} P_d(t, T_k) \mathbb{E}^{Q_d^{T_k}}[\pi_k | \mathcal{F}_t].$$ |

**Table 1:** comparison table between the classical Single-Curve methodology and the modern Multiple-Curve methodology. We refer to a general single-currency interest rate derivative under CSA characterized by $m$ future coupons with payoffs $\boldsymbol{\pi} = \{\boldsymbol{\pi_1}, \ldots, \boldsymbol{\pi_m}\}$, generating $m$ cash flows $\boldsymbol{c} = \{\boldsymbol{c_1}, \ldots, \boldsymbol{c_m}\}$ at future dates $\mathrm{T} = \{\mathrm{T_1}, \ldots, \mathrm{T_m}\}$, with $\boldsymbol{t} < \boldsymbol{T_1} < \cdots < \boldsymbol{T_m}$.

|  | **Classical Approach (Single-Curve)** | **Modern Approach (Multiple-Curve)** |
|---|---|---|
| **FRA** | $V_{FRA}(t;T_{k-1},T_k,K,\omega) = NP(t,T_k)\omega[F_k(t) - K]\tau_k,$<br><br>with<br><br>$F_k(t) := F(t;T_{k-1},T_k) = \mathbb{E}^{Q^{T_k}}[L(T_{k-1},T_k)|\mathcal{F}_t].$ | $V_{FRA}(t;T_{k-1},T_k,K,\omega) = NP_d(t,T_k)\omega[\tilde{F}_{x,k}(t) - K]\tau_{x,k},$<br><br>with<br><br>$\tilde{F}_{x,k}(t) := \tilde{F}_x(t;T_{k-1},T_k) = \mathbb{E}^{Q_d^{T_k}}[L_x(T_{k-1},T_k)|\mathcal{F}_t].$ |
| **Swap** | $Swap(t,\mathbf{T},\mathbf{S},K,\omega)$<br>$= N\omega\left[\sum_{k=1}^{m} P(t,T_k)F_k(t)\tau_k - KA(t,\mathbf{S})\right],$<br><br>with<br><br>$K = R^{Swap}(t;\mathbf{T},\mathbf{S}) = \frac{\sum_{k=1}^{m} P(t,T_k)F_k(t)\tau_k}{A(t,\mathbf{S})},$<br><br>$A(t,\mathbf{S}) = \sum_{j=1}^{n} P(t,S_j)\tau_j.$ | $Swap(t,\mathbf{T},\mathbf{S},K,\omega)$<br>$= N\omega\left[\sum_{k=1}^{m} P_d(t,T_k)\tilde{F}_{x,k}(t)\tau_{x,k} - KA_d(t,\mathbf{S})\right],$<br><br>with<br><br>$K = \tilde{R}_x^{Swap}(t;\mathbf{T},\mathbf{S}) = \frac{\sum_{k=1}^{m} P_d(t,T_k)\tilde{F}_{x,k}(t)\tau_{x,k}}{A_d(t,\mathbf{S})},$<br><br>$A_d(t,\mathbf{S}) = \sum_{j=1}^{n} P_d(t,T_j)\tau_j.$ |
| **Basis Swap** | $BS(t;\mathbf{T}_x,\mathbf{T}_y,Z,\omega)$<br>$= N\omega\left[\sum_{j=1}^{m_x} P(t,T_{x,j})\left(F_{x,j}(t) + Z(t;\mathbf{T}_x,\mathbf{T}_y)\right)\tau_{x,j} - \sum_{k=1}^{m_y} P(t,T_{y,k})F_{y,k}(t)\tau_{y,k}\right],$<br><br>with<br><br>$Z(t;\mathbf{T}_x,\mathbf{T}_y)$<br>$= \frac{\sum_{k=1}^{m_y} P(t,T_{y,k})F_{y,k}(t)\tau_{y,k} - \sum_{j=1}^{m_x} P(t,T_{x,j})F_{x,j}(t)\tau_{x,j}}{A(t,\mathbf{T}_x)}$<br>$= 0.$ | $BS(t;\mathbf{T}_x,\mathbf{T}_y,Z,\omega)$<br>$= N\omega\left[\sum_{j=1}^{m_x} P_d(t,T_{x,j})\left(\tilde{F}_{x,j}(t) + Z(t;\mathbf{T}_x,\mathbf{T}_y)\right)\tau_{x,j} - \sum_{k=1}^{m_y} P_d(t,T_{y,k})\tilde{F}_{y,k}(t)\tau_{y,k}\right],$<br><br>with<br><br>$Z(t;\mathbf{T}_x,\mathbf{T}_y)$<br>$= \frac{\sum_{k=1}^{m_y} P_d(t,T_{y,k})\tilde{F}_{y,k}(t)\tau_{y,k} - \sum_{j=1}^{m_x} P_d(t,T_{x,j})\tilde{F}_{x,j}(t)\tau_{x,j}}{A_d(t,\mathbf{T}_x)}.$ |
| **Cap/Floor** | $CF^{Black}(t;\mathbf{T},K,\omega) = \sum_{k=2}^{m} cf(t;T_{k-1},T_k,K,\omega),$<br><br>with<br><br>$cf(t;T_{k-1},T_k,K,\omega)$<br>$= N\omega P(t,T_k)[F_k(t)\phi(\omega d_k^+) - K\phi(\omega d_k^-)]\tau_k,$<br><br>$d_k^\pm = \frac{\ln\left(\frac{F_k(t)}{K}\right) \pm \frac{1}{2}\sigma_k(t)^2\tau_{k-1}(t)}{\sigma_k(t)\sqrt{\tau_{k-1}(t)}}.$ | $CF^{Black}(t;\mathbf{T},K,\omega) = \sum_{k=2}^{m} cf(t;T_{k-1},T_k,K,\omega),$<br><br>with<br><br>$cf(t;T_{k-1},T_k,K,\omega)$<br>$= N\omega P_d(t,T_k)[\tilde{F}_{x,k}(t)\phi(\omega d_k^+) - K\phi(\omega d_k^-)]\tau_{x,k},$<br><br>$d_k^\pm = \frac{\ln\left(\frac{\tilde{F}_{x,k}(t)}{K}\right) \pm \frac{1}{2}\tilde{\sigma}_{x,k}(t)^2\tau_{x,k-1}(t)}{\tilde{\sigma}_{x,k}(t)\sqrt{\tau_{x,k-1}(t)}}.$ |

**Table 2:** comparison table of classical and modern formulas for pricing plain vanilla derivatives. In red we emphasize the most relevant peculiarities of the Multiple-Curve method.

# 4. Empirical Pricing Analysis

In the following sections we present the results of an empirical analysis comparing the results of the three pricing frameworks described before against market quotations of plain vanilla interest rate derivatives at two different valuation dates. The aim of this analysis is to highlight the time evolution of the market pricing approach as a consequence of the financial crisis.

## 4.1. Market Data, Volatility Surfaces and Yield Curves

The reference market quotes that we considered are, in particular:

- **Euro Forward Start Interest Rate Swap** contracts (FSIRS): market swap rates based on Euribor 6M, published by Reuters.

- **Euro Cap/Floor** options: market premia and Black implied volatility surfaces based on Euribor 6M, published by Reuters.

All the market data were archived at close of business (17:30 CET) on 31$^{st}$ March and 31$^{st}$ August 2010, and refer to instruments traded among collateralized counterparties in the European interbank market.

Moreover, for the pricing analysis we defined the following yield curves:

- **Euribor Standard**: the classical yield curve bootstrapped from short term Euro Deposits (from 1D to the first front Futures), mid term Futures on Euribor 3M (below 3Y) and mid-long term Swaps on Euribor 6M (after 3Y) (see Table 1, left column).

- **Euribor 6M Standard**: the "pure" Euribor 6M forwarding yield curve is bootstrapped from the Euro Deposit 6M, mid term FRAs on Euribor 6M (below 3Y) and mid-long term Swaps on Euribor 6M (after 3Y). The discounting curve used in the bootstrapping procedure is the Euribor Standard. This curve differs from the one above on the short-mid term, while after the 3-year node it tends to coincide with the Euribor Standard since the bootstrapping instruments are the same.

- **Eonia OIS**: the modern discounting yield curve bootstrapped from quoted Eonia OIS. This yield curve usually presents lower rates and, hence, greater discount factors than those obtained with the Euribor Standard above.

- **Euribor 6M CSA**: the modern Euribor 6M forwarding yield curve bootstrapped from the same market instruments as the Euribor 6M Standard curve, but using the Eonia OIS as discounting curve. This curve usually presents small , but not negligible, differences with respect to the Euribor 6M Standard.

In Figure 6 and Figure 7 we plot the yield curves introduced above and the Forward/FRA rates considered in the analysis. The data refers to the 31$^{st}$ March 2010. The curves at 31$^{st}$ August 2010 are similar.

Figure 6 shows the term structure of the Euribor Standard, of the Euribor 6M Standard, of the Euribor 6M CSA and of the Eonia OIS zero rate curves. The three Euribor-curves present a very similar term structure. The 6-month tenor curves (Euribor 6M Standard and Euribor 6M CSA), show differences below 2 bps. The Euribor Standard curve differs from the other two curves in the short term window, since up to 3-year maturity it is bootstrapped using market Futures on Euribor 3M. The fourth yield curve in Figure 6, Eonia OIS, is different (lower than Euribor curves), including lower credit and liquidity risk components that differentiates the risky rates form the risk free ones.

In Figure 7 we report the three different forwarding curves we used in the analysis: the Euribor Standard 6-month forward rate curve, the Euribor 6M Standard 6-month FRA rate curve and the Euribor 6M CSA 6-month FRA rate curve. Even if the differences between these three curves are lower than 2 bps, we will observe that the choice of the correct forwarding curve has a relevant impact in the pricing of interest rate derivatives, especially for the longer maturities.

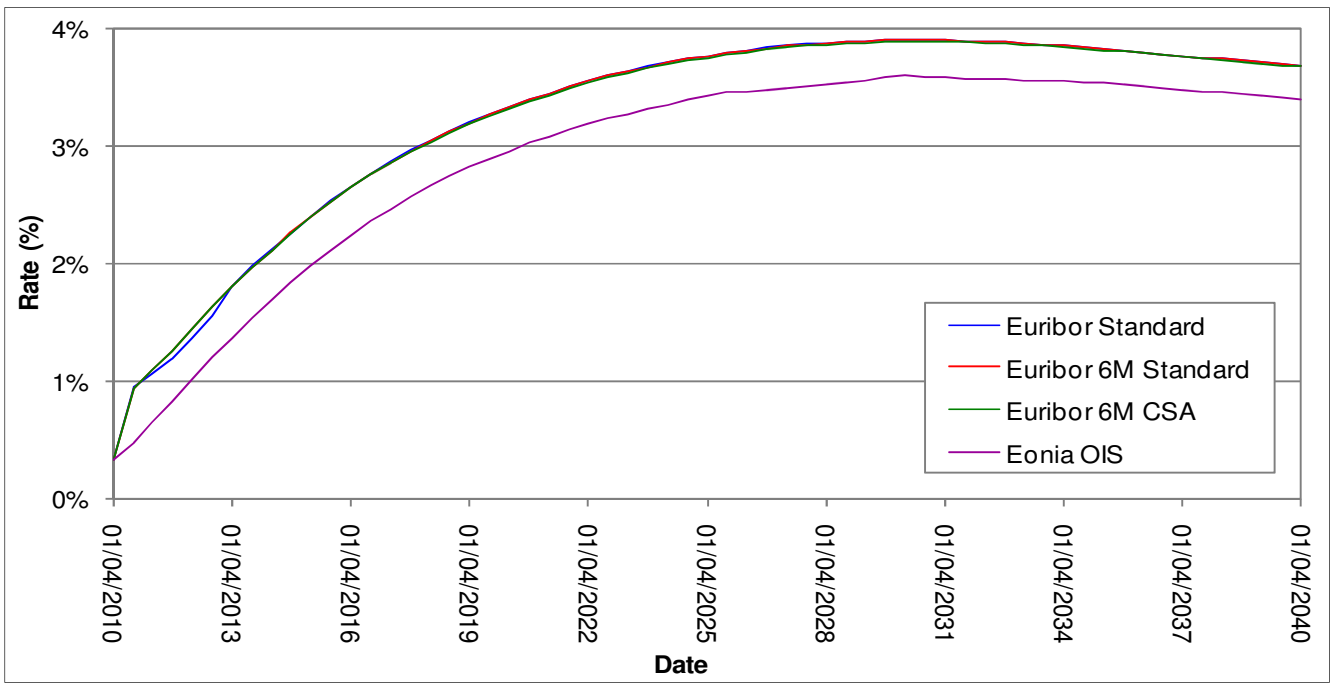

**Figure 6:** term structures of the Euribor Standard, the Euribor 6M Standard, the Euribor 6M CSA and Eonia OIS zero rate curves. Reference date: 31$^{st}$ March 2010 (source: Reuters).

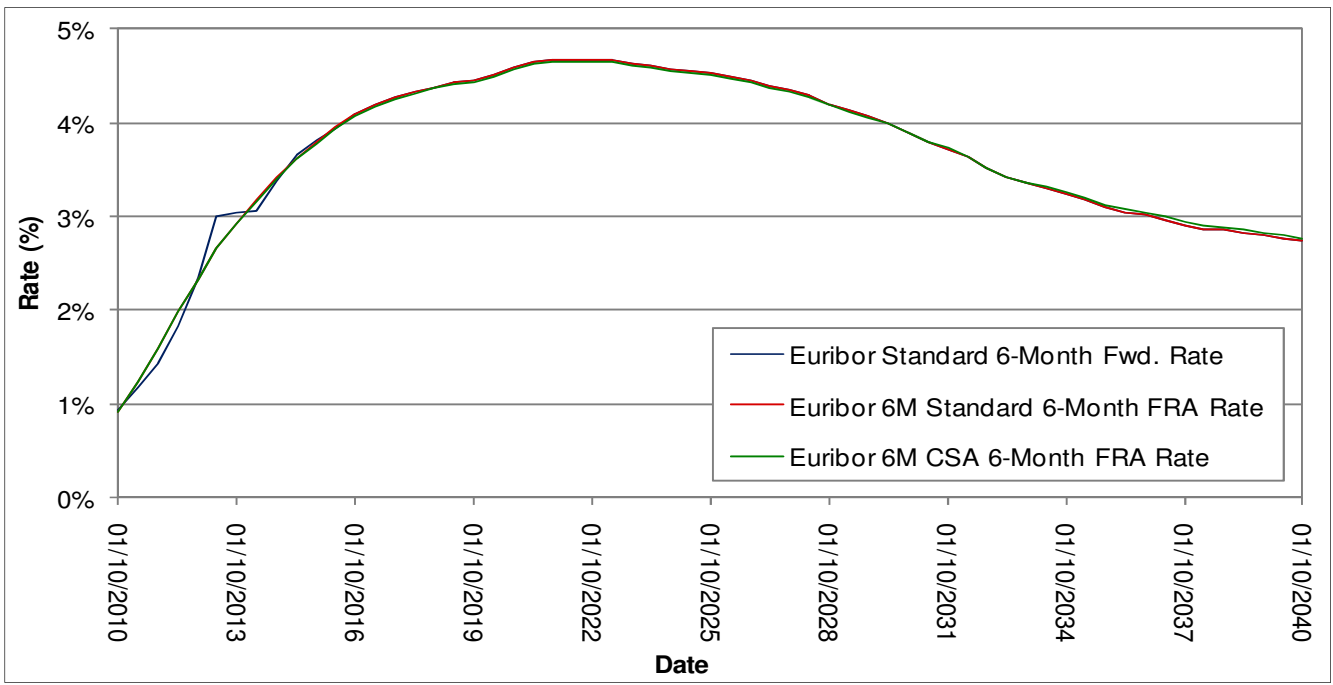

**Figure 7:** term structures of the Euribor Standard 6-month forward rate, of the Euribor 6M Standard 6-month FRA rate and of the Euribor 6M CSA 6-month FRA rate. Note that the Euribor Standard 6-month forward rate coincides with the corresponding 6-month FRA rate since the yield curve is consistent with the Single-Curve framework. Reference date: 31$^{st}$ March 2010 (source: Reuters).

## 4.2. Pricing Methodologies

We have tested three different pricing methodologies as described below.

1.  **Single-Curve approach**: we use the Euribor Standard yield curve to calculate both the discount factors $P(t, T_k)$ and the forward rates $F$ needed for pricing any interest rate derivatives. This is the classical Single-Curve methodology adopted by the market before the credit crunch, without collateral, credit and liquidity effects.

2.  **Multiple-Curve No-CSA approach**: we calculate discount factors $P_d(t, T_k)$ on the Euribor Standard curve and FRA rates $\tilde{F}_{x,k}(t)$ on the Euribor 6M Standard curve. This is the "quick and dirty" methodology adopted by the market in response to the credit crunch after August 2007, that distinguishes between discounting and forwarding curves. It is defined "No-CSA" because it does not include the effect of collateral. Indeed, the Euribor Standard discounting curve reflects the average cost of (uncollateralized) funding of a generic European interbank counterparty (belonging to the Euribor panel). Also the Euribor 6M Standard forwarding curve construction does not take into account collateralization, but it does include the tenor-specific credit and liquidity risk of the underlying Euribor 6M rate.

3.  **Multiple-Curve CSA approach**: we calculate discount factors $P_d(t, T_k)$ on the Eonia OIS curve and FRA rates $\tilde{F}_{x,k}(t)$ on the Euribor 6M CSA curve. This is the "state of the art" modern methodology, fully coherent with the CSA nature of the interest rate derivatives considered and with the credit and liquidity risk of the underlying Euribor 6M rate.

The arbitrage-free formulas used in the analysis are those reported in Table 2: the Single-Curve approach is in the left column) and the two Multiple-Curve approaches are on the right column. The two following sections report the findings of the analysis.

## 4.3. Forward Start Interest Rate Swaps

The Forward Start Interest Rate Swap contracts considered here are characterized by a floating leg on Euribor 6M with 6-month frequency vs a fixed leg with annual frequency, a forward start date and maturity dates ranging from 1 to 25 years. We selected forward start instead of spot start swaps since the former are more sensible to the choice of the pricing methodology. For each methodology (section 4.2) and each valuation date (31st March and 31st August 2010) we computed the theoretical equilibrium FSIRS rates and we compared them with the market quotes. The results are shown in Figure 8 below, while in Table 3 we compare the most important numbers: the range of minimum and maximum discrepancies and the standard deviation. The spikes observed in the figures for start dates below 3Y may be explained in terms of differences in the short term yield curve construction, where there is a significant degree of freedom in choosing the bootstrapping instruments (Deposits, FRAs and Futures). Smaller spikes are also present for short tenor FSIRS with maturity below 3Y because these swaps depend on a few forwards and discounts and, thus, are more sensitive to minor differences in the yield curves. Hence we show results in Table 3 both including and excluding the two "stripes" below 2 years start/maturity date considering both the valuation dates (31st March and 31st August 2010).

From Figure 8 we observe that the first methodology has the worst performance, producing, on average, overestimated FSIRS rates. The second methodology introduces small improvements, at least below 3 years. This is expected, because the two curves used are very similar after 3 years, both using standard Euro Swaps on Euribor 6M. The third methodology is by far the best in reproducing the market data. The remaining differences around 1 basis points may be explained with minor differences with respect to the market yield curves. The same observations apply to the results of 31st August 2010.

We conclude that the market of Interest Rate Swaps since March 2010 has abandoned the classical Single-Curve pricing methodology, typical of the pre-credit crunch interest rate world, and has adopted the modern Multiple-Curve CSA approach, thus incorporating into market prices the credit and liquidity effects described in section 4.2 above.

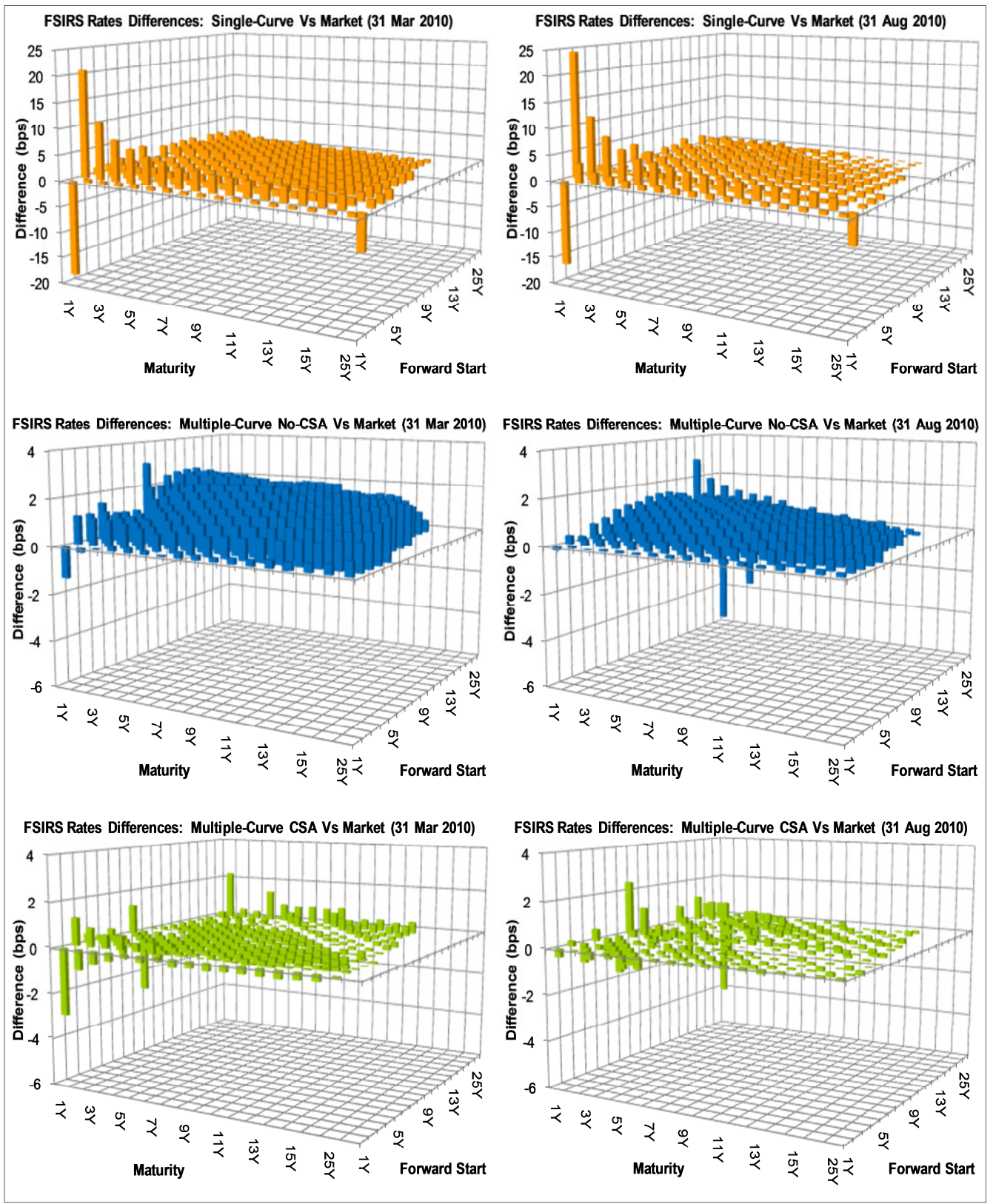

**Figure 8**: FSIRS rates differences. Upper panels: differences between theoretical Single-Curve FSIRS rates and market quotes. Middle panels: differences between theoretical Multiple-Curve No-CSA FSIRS rates and market quotes. Lower panels: differences between theoretical Multiple-Curve CSA FSIRS rates and market quotes. Valuation dates: 31$^{st}$ March 2010 (left side graphs) and 31$^{st}$ August 2010 (right side graphs). Note that the y-axis scale of the middle and lower graphs has been magnified in order to better appreciate lower price differences (source: Reuters).

| **Forward Start Interest Rate Swaps Differences** | | | | | | | | |
|---|---|---|---|---|---|---|---|---|
| | **31st March 2010** | | | | **31st August 2010** | | | |
| | **Range** | | **Standard deviation** | | **Range** | | **Standard deviation** | |
| **Single-Curve** | [-18.4;+20.8] | [-3.2;+2.7] | 2.84 | 1.89 | [-16.3;+24.4] | [-3.9;+1.9] | 2.58 | 1.15 |
| **Multiple-Curve No-CSA** | [-2.9;+3.1] | [-2.9;+2.6] | 1.77 | 1.86 | [-5.7;+2.9] | [-3.7;+1.7] | 1.11 | 1.09 |
| **Multiple-Curve CSA** | [-2.9;+2.3] | [-1.0;+1.5] | 0.53 | 0.37 | [-4.1;+2.4] | [-1.4;+1.0] | 0.47 | 0.26 |

**Table 3:** differences (in basis points) from Figure 8. For each pricing methodology (section 4.2) and each valuation date (31st March and 31st August 2010) we show the range of minimum and maximum discrepancies and the standard deviation, both considering all FSIRS (columns on the left) and excluding the two 1Y-2Y stripes (columns on the right).

### 4.4. Cap/Floor Options

The European Cap/Floor options considered here are characterized by floating payments with 6M frequency indexed to Euribor 6M, spot start date, maturity dates ranging from 3 to 30 years, and strikes ranging from 1% to 10%. The first caplet/floorlet, already known at spot date, is not included in the cap/floor premium. The market quotes Floor premia for strikes below the at-the-money (ATM) and Cap premia for strikes above ATM. For each methodology (section 4.2) and each valuation date (31st March and 31st August 2010) we computed the theoretical Caps/Floors premia and we compared them with the market premia.

The computation of the theoretical European Caps/Floors premia using the standard Black's formula in Table 2 requires two inputs: the pair of discounting and forwarding curves and the Black implied term volatility (see e.g. Mercurio (2009, 2010b)). Even if the market-driven quantity is the premium of the traded option, it's standard market convention to quote the option in terms of its Black implied term volatility. Clearly, once the premium is fixed by the market supply and demand, the value of this volatility depends on the curves used for discounting and forwarding. Thus, a change in the market yield curves implies a corresponding change in the Black implied volatilities. Actually, in August 2010 the market begun to quote two distinct volatility surfaces: one implied using the classical Euribor discounting curve and one implied using the modern Eonia discounting curve (see ICAP (2010)). The Eonia implied volatility is generally lower than the Euribor implied volatility, because the effect of lower Eonia rates (higher Eonia discount factors) must be compensated with lower values of implied volatility. Coherently with the market change, we used the Euribor volatility to compute Multiple-Curve No-CSA prices and the Eonia volatility to compute Multiple-Curve CSA prices at 31st August 2010. The results are shown in Figure 9 and all the relevant numbers are contained in Table 4.

Overall, we notice again that, on both dates, the Single-Curve methodology (upper panels) has a very bad performance. The Multiple-Curve No-CSA methodology (middle panels) has a good performance on both dates, with an absolute average difference of 1.4/1.6 bps over a total of 169 options and a standard deviation of 2.06/2.28 bps. Finally the Multiple-Curve CSA methodology (lower panels) shows a bad performance on the first date (standard deviation 15.82 bps) and a performance as good as that of the Multiple-Curve CSA methodology on the second date, with absolute average difference of 1.7 bps and standard deviation of 2.43 bps.

We conclude that the results discussed above are coherent with the interest rate market evolution after the credit crunch and, in particular, with the market changes announced in August 2010 (ICAP (2010)) and with our findings for Forward Start IRS discussed in section 4.3. First of all, the market, at least since March 2010, has abandoned the classical Single-Curve pricing methodology, typical of the pre-credit crunch interest rate world, and has adopted the modern Multiple-Curve approach. Second, the transition to the CSA-discounting methodology for options has happened just in August 2010, thus incorporating into market prices the credit and liquidity effects described in section 4.2 above. In the latter case, contrary to FSIRS, both the two modern Multiple-Curve methodologies (if correctly applied) lead to good repricing of the market premia, because, at constant market premia, the change in the yield curves (switching from Euribor discounting to Eonia discounting) are compensated by the corresponding changes in the Black implied volatilities.

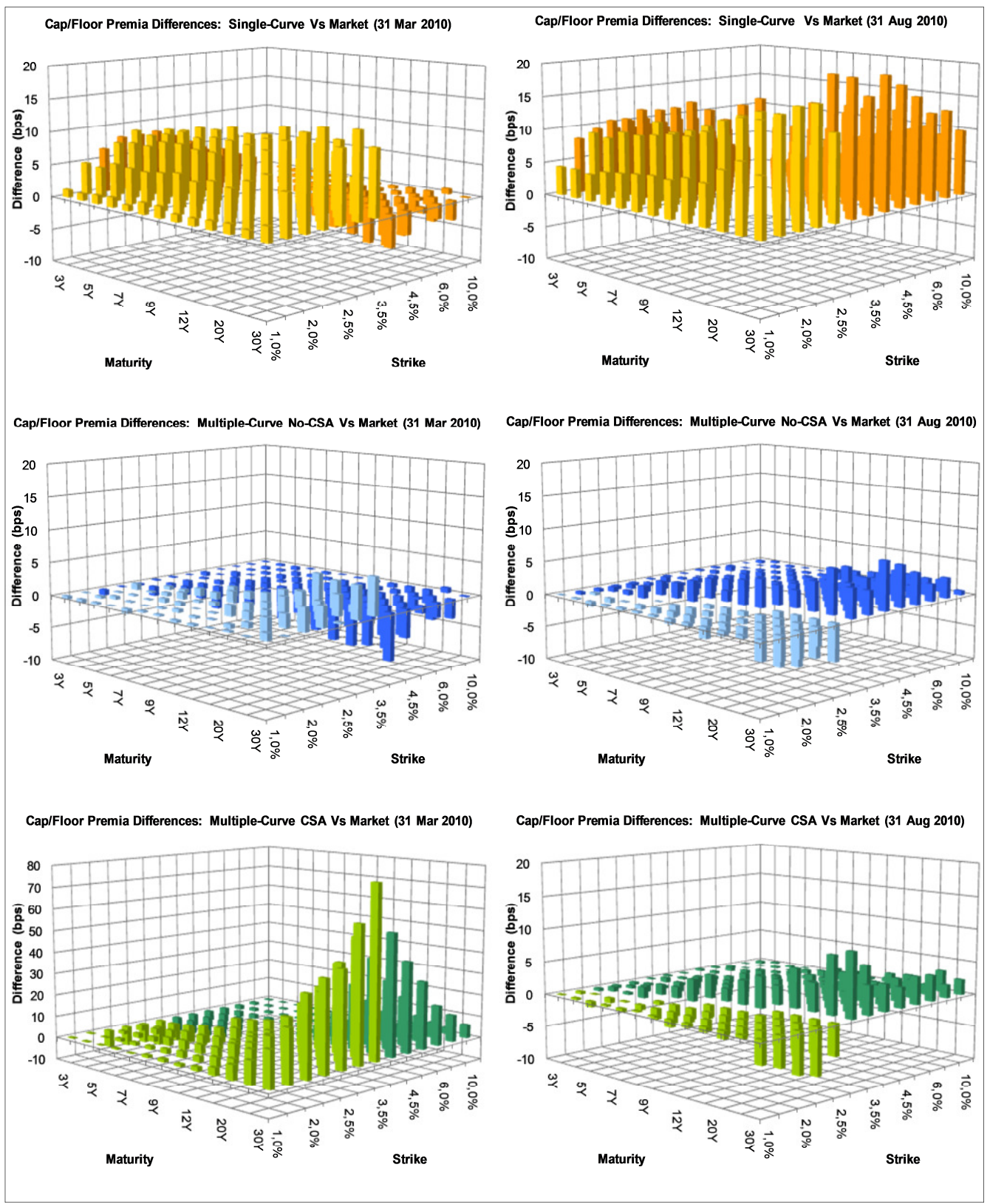

**Figure 9:** cap/floor options premia differences (light colours: Floors, dark colours: Caps). Upper panels: differences between Single-Curve premia and market premia. Middle panels: differences between Multiple-Curve No-CSA premia and market premia. Lower panels: differences between Multiple-Curve CSA premia and market premia. Valuation dates: 31$^{st}$ March 2010 (left side graphs) and 31$^{st}$ August 2010 (right side graphs). Note that the y-axis scale of the lower graph on the left side has been reduced to better highlight larger price differences (source: Reuters).

| Cap/Floor Premia Differences | | | | |
|---|---|---|---|---|
| | 31st March 2010 | | 31st August 2010 | |
| | Range | Standard deviation | Range | Standard deviation |
| **Single-Curve** | [-5.8;+14.1] | 6.3 | [+0.2;+20.0] | 9.7 |
| **Multiple-Curve No-CSA** | [-7.0;+5.8] | 2.1 | [-6.3;+7.4] | 2.3 |
| **Multiple-Curve CSA** | [-8.9;+77.7] | 15.8 | [-6.8;+9.6] | 2.4 |

**Table 4:** differences (in basis points) from Figure 9. For each pricing methodology (section 4.2) and each valuation date (31st March and 31st August 2010) we show the range of minimum and maximum discrepancies and the standard deviation.

## 4.5. SABR Model Calibration

The SABR (Stochastic Alpha Beta Rho) model developed by Hagan et al. (2002) is one of the simplest generalizations of the well known Black's model with stochastic volatility, preserving Black-like closed formulas for Caps, Floors and Swaptions, leading to a market coherent description of the dynamics of the volatility and allowing calibration to the interest rate smile. Thanks to its mathematical simplicity and transparent financial interpretation, it has established itself as the market standard for pricing and hedging plain vanilla interest rate options and to calibrate the market volatility, often called "SABR volatility surface" (for Caps/Floors) or "cube" (for Swaptions).

| | **Classical SABR (Single-Curve)** | **Modern SABR (Multiple-Curve)** |
|---|---|---|
| **SABR dynamics** | $\frac{dF_k(t)}{F_k^\beta(t)} = V(t)dW_k^{Q^{T_k}}(t),$ $\frac{dV(t)}{V(t)} = \nu dZ^{Q^{T_k}}(t),$ $V(t_0) = \alpha,$ $dW_k^{Q^{T_k}}(t)dZ^{Q^{T_k}}(t) = \rho dt,$ for each $k = 1, \dots, m,$ $t < T_{k-1} < T_k, \nu > 0, \alpha > 0, 0 \leq \beta \leq 1, -1 \leq \rho \leq 1.$ | $\frac{d\tilde{F}_{x,k}(t)}{\tilde{F}_{x,k}^\beta(t)} = V_x(t)dW_{x,k}^{Q_d^{T_k}}(t),$ $\frac{dV_x(t)}{V_x(t)} = \nu dZ_x^{Q_d^{T_k}}(t),$ $V_x(t_0) = \alpha,$ $dW_{x,k}^{Q_d^{T_k}}(t)dZ_x^{Q_d^{T_k}}(t) = \rho dt,$ for each $k = 1, \dots, m,$ $t < T_{k-1} < T_k, \nu > 0, \alpha > 0, 0 \leq \beta \leq 1, -1 \leq \rho \leq 1$ |
| **SABR volatility** | $\tilde{\sigma}_x^{SABR}(t; T_{k-1}, T_k, K) = \frac{\nu \ln\frac{\tilde{F}_{x,k}(t)}{K}}{x(z)} \frac{A(\tilde{F}_{x,k}(t),K)}{B(\tilde{F}_{x,k}(t),K)}$ $A(\tilde{F}_{x,k}(t), K) := 1 + \left[\frac{\alpha^2(1-\beta)^2}{24(\tilde{F}_{x,k}(t)K)^{1-\beta}} + \frac{\alpha\beta\nu\rho}{4(\tilde{F}_{x,k}(t)K)^{\frac{(1-\beta)}{2}}} + \nu^2\frac{2-3\rho^2}{24}\right]\tau_{x,k} + \cdots,$ $B(\tilde{F}_{x,k}(t), K) := 1 + \frac{1}{24}\left[(1-\beta)\ln\frac{\tilde{F}_{x,k}(t)}{K}\right]^2 + \frac{1}{1920}\left[(1-\beta)\ln\frac{\tilde{F}_{x,k}(t)}{K}\right]^4 + \cdots,$ $x(z) := \ln\frac{\sqrt{1-2\rho z + z^2} + z - \rho}{1-\rho}, \quad z := \frac{\nu}{\alpha}(\tilde{F}_{x,k}(t)K)^{\frac{(1-\beta)}{2}}\ln\frac{\tilde{F}_{x,k}(t)}{K}.$ ||

**Table 5:** classical (top left column) vs modern (top right column) SABR model dynamics and volatility expression consistent with the Multiple-Curve approach (bottom). See Hagan et al. (2002) for details.

In analogy with the Black model, the modern version of the SABR model is obtained from the corresponding classical SABR version of Hagan et al. (2002) just by replacing the classical forward rate with the modern FRA rate and the $T_K$-forward Libor measure associated with the classical Single-Curve numeraire $P(t,T_k)$ with the modern $T_K$-forward measure associated with the discounting numeraire $P_d(t,T_k)$. The SABR volatility formula remains unchanged, but takes the FRA rate in input. Caps/Floor options are priced as in Table 2 using the standard Black's formula and input SABR volatility. In Table 5 we show the classical and the modern SABR equations.

Using different Multiple-Curve pricing methodologies as in section 4.2, based on different choices of discounting and forwarding yield curves, leads to the definition of two distinct implied volatility surfaces referring to the same collateralized market premia, as discussed in section 4.4 above:

- the Euribor implied term volatility, which is consistent with the Multiple-Curve No-CSA approach;
- the Eonia implied term volatility, which is consistent with the Multiple-Curve CSA approach.

Notice that the SABR model refers to forward (not term) volatilities implied in caplets/floorlets (not Caps/Floors). We denote with $\tilde{\sigma}_x(t; T_{k-1}, T_k, K)$ the implied forward volatility seen at time $t$ of an European caplet/floorlet on the spot Euribor rate $L_x(T_{k-1}, T_k)$ and strike $K$, with x = {Euribor 6M Standard, Euribor 6M CSA}. Thus, we stripped the two forward volatility surfaces implied in the Cap/Floor premia published by Reuters on the 31st March and on 31st August 2010, using the two Multiple-Curve methodologies above. The stripping procedure requires many technicalities that we do not report here, we refer to section 3.6 in Brigo and Mercurio (2006).

The SABR calibration procedure is applied to each smile section, corresponding to the strip of caplets/floorlets with the same maturity date $T_k$, underlying FRA rate $\tilde{F}_{x,k}(t)$, and different strikes $K_j$, $j = \{1, ..., 14\}$[3]. Thus the calibration returns the values of the model's parameters $\alpha, \beta, \nu, \rho$ that minimize the distance between the market implied forward volatilities $\tilde{\sigma}_x^{Mkt}(t; T_{k-1}, T_k, K_j)$ and the corresponding theoretical SABR volatilities $\tilde{\sigma}_x^{SABR}(t; T_{k-1}, T_k, K_j)$ obtained through the closed analytic formula in Table 5. Thus we obtain a set of SABR parameters for each smile section.

For the two dates (31st March and on 31st August 2010) and the two pricing methodologies (Multiple-Curve No-CSA, Multiple-Curve CSA) associated to the two corresponding forward volatility surfaces (Euribor, Eonia), we performed two minimizations using two distinct error functions:

- a standard error function defined as the square root of the sum of the square differences between the SABR and the market forward volatilities:

$$Error_{std}(T_k) = \left\{ \sum_{j=1}^{14} \left[ \tilde{\sigma}_x^{Mkt}(t; T_{k-1}, T_k, K_j) - \tilde{\sigma}_x^{SABR}(t; T_{k-1}, T_k, K_j) \right]^2 \right\}^{\frac{1}{2}} \quad [1]$$

- a vega-weighted error function:

$$Error_{vw}(T_k) = \left\{ \sum_{j=1}^{14} \left[ \left( \tilde{\sigma}_x^{Mkt}(t; T_{k-1}, T_k, K_j) - \tilde{\sigma}_x^{SABR}(t; T_{k-1}, T_k, K_j) \right) W_{j,x} \right]^2 \right\}^{\frac{1}{2}} \quad [2]$$

where

$$W_{j,x} = \frac{\upsilon(T_k, K_j)}{\sum_{j=1}^{14} \upsilon(T_k, K_j)}$$

and $\upsilon(T_k, K_j)$ is the Black's vega sensitivity of the caplet/floorlet option with strike $K_j$, FRA rate $\tilde{F}_{x,k}(t)$ and maturity $T_k$. Weighting the errors by the sensitivity of the options to shifts of the volatility allows, during the calibration procedure, to give more importance to the near-ATM areas of the volatility surface, with high vega sensitivities and market liquidity, and less importance to OTM areas, with lower vega and liquidity.

The initial values of $\alpha, \rho,$ and $\nu$ were respectively 4.5%, -10% and 20%. Different initializations gave no appreciable differences in the calibration results. According to Hagan et al. (2002) and West (2004), in the calibration of the model we decided to fix the value of the redundant parameter $\beta$ to 0.5. The minimization was performed using the built-in Matlab's function "patternsearch".

A snapshot of the SABR calibration results is shown in Figure 10 and Figure 11, where we report three smile sections at short term (2-year maturity), mid term (10-year maturity) and long term (30-year maturity). In Figure 12 (valuation date: 31st March 2010) and in Figure 13 (valuation date: 31st August 2010) we plot the vega-weighted calibration errors we obtain using the two different minimization functions. For each smile section, the errors of the standard calibration are equally weighted, while those obtained through the vega-weighted approach are weighted for the volatility sensitivity of caplet/floorlet options given a certain strike and maturity date. In Table 6 we compare the two calibration approaches reporting the most important numbers: the range of minimum and maximum errors and the standard deviation.

---

[3] 14 is the number of strikes quoted in the market.

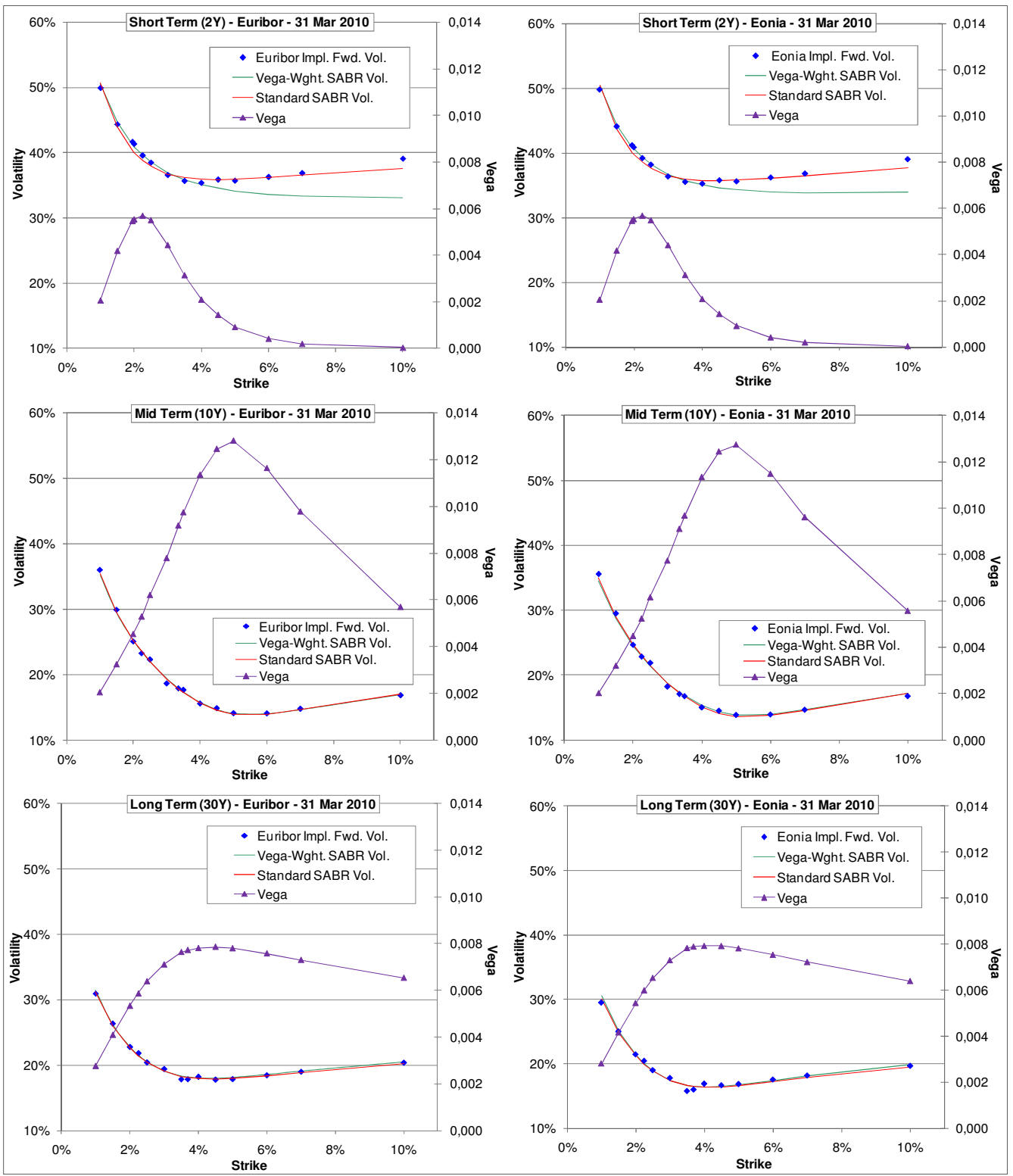

**Figure 10:** SABR model calibration results. The blue dots represents the market implied forward volatility, the red line refers to the standard calibration, the green line refers to the vega-weighted calibration and the purple line (right y-axis) report the values of the vega. The graphs on the left are related to the market Euribor implied forward volatility. The graphs on the right are associated to the market Eonia implied forward volatility. Upper panels: smile section with maturity date 2-year. Middle panels: smile section with maturity date 10-year. Lower panels: smile section with maturity date 30-years. Valuation date: 31$^{st}$ March 2010 (source: Reuters).

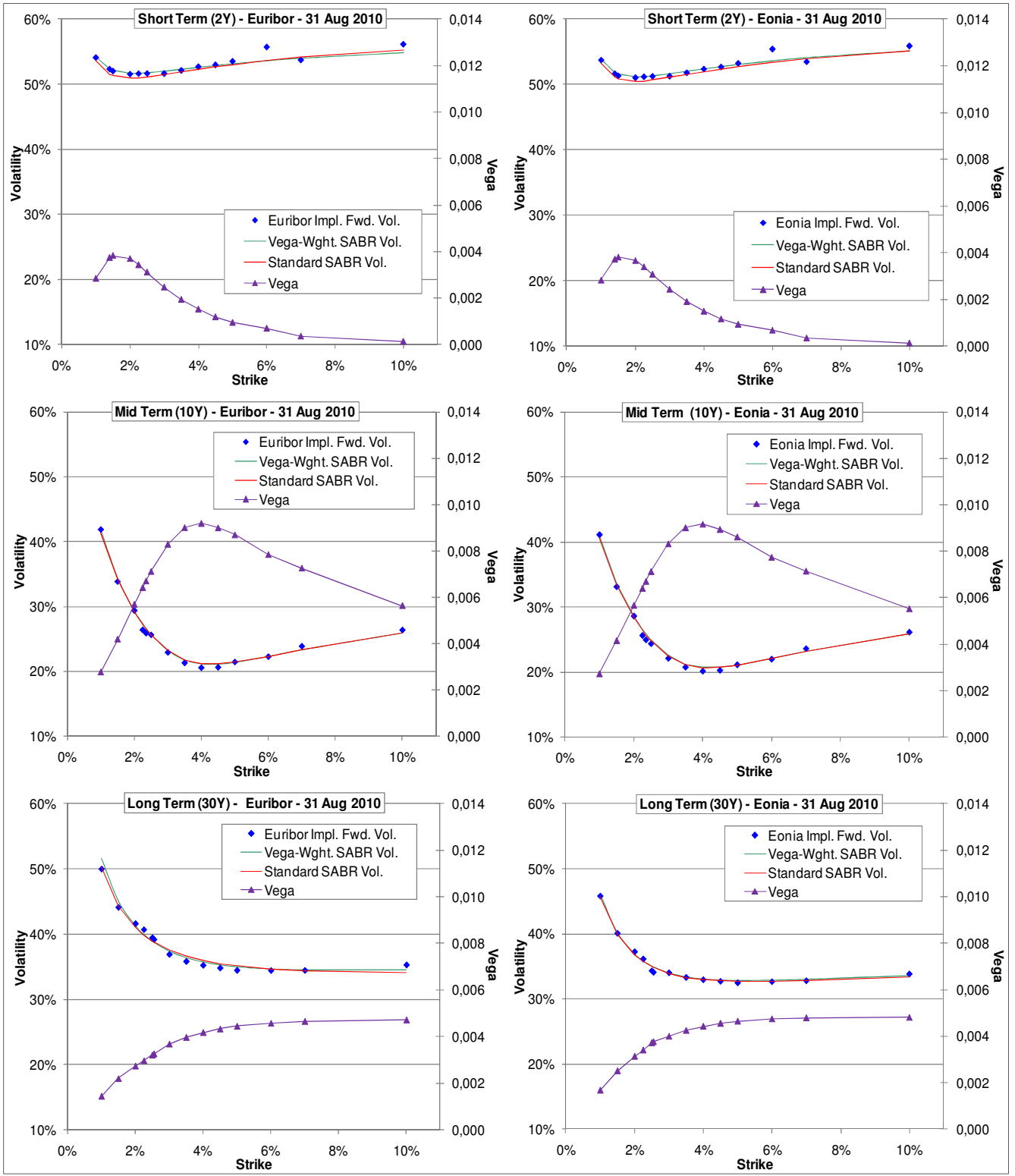

**Figure 11:** same as Figure 10, valuation date: 31$^{st}$ August 2010.

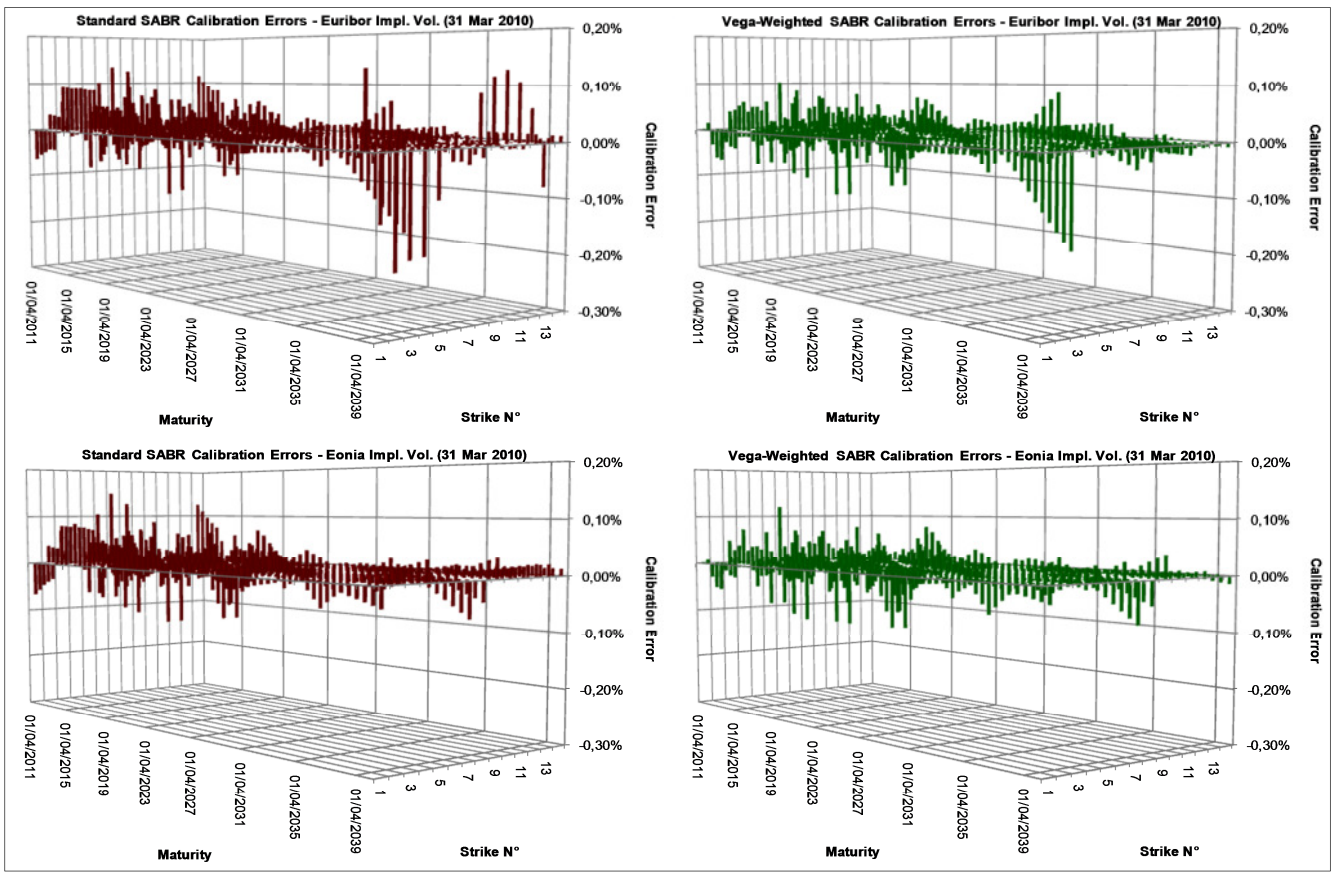

**Figure 12:** SABR calibration errors. Upper/lower panels: SABR calibration on the Euribor/Eonia implied volatility surface. Left/right panels: standard/vega-weighted SABR calibration. Valuation date: 31$^{st}$ March 2010 (source: Reuters).

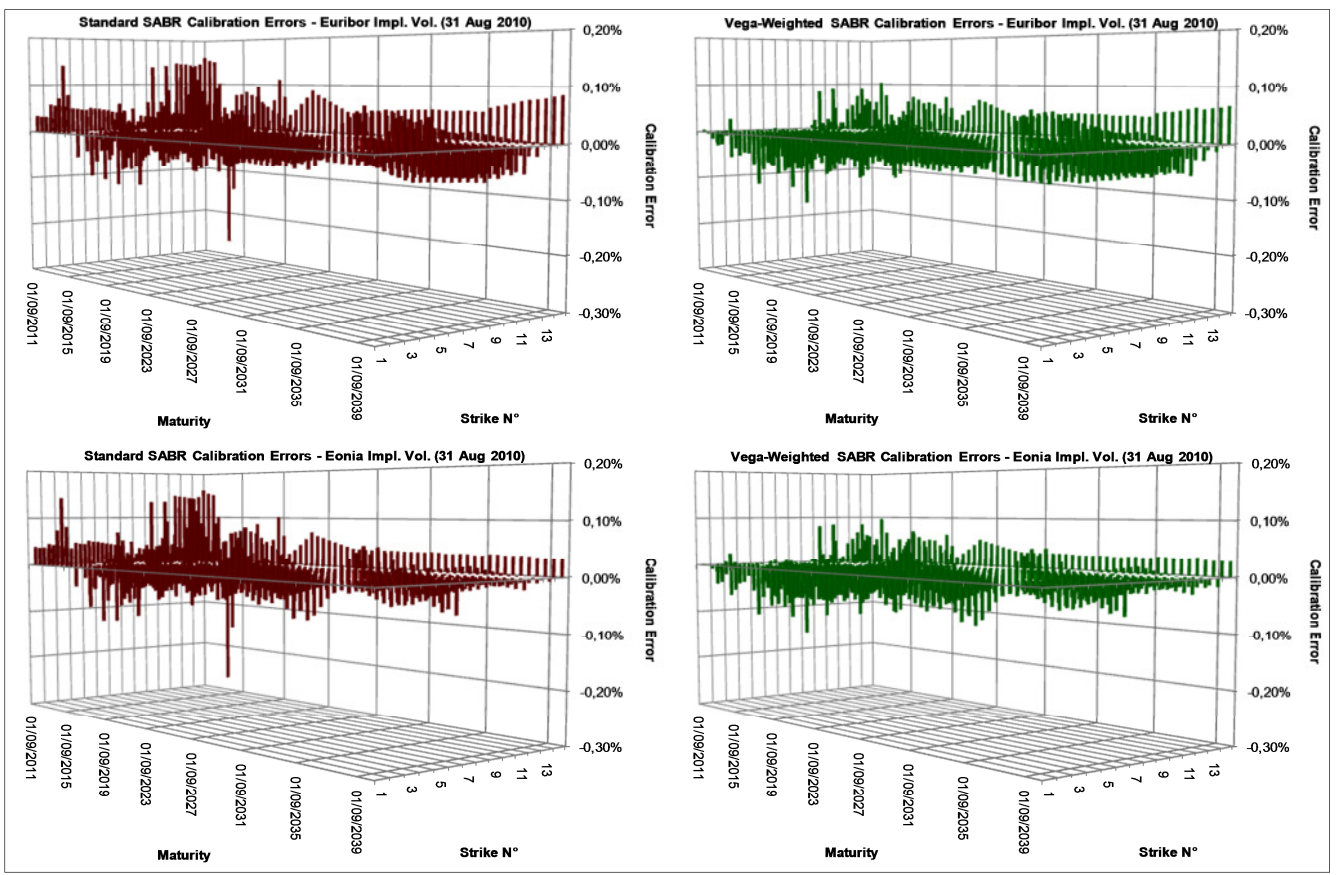

**Figure 13:** same as Figure 12, valuation date: 31$^{st}$ August 2010.

| | SABR Calibration Errors | | | | | | | |
|---|---|---|---|---|---|---|---|---|
| | 31st March 2010 | | | | 31st August 2010 | | | |
| | Implied Volatility Euribor | | Implied Volatility Eonia | | Implied Volatility Euribor | | Implied Volatility Eonia | |
| | Range | Standard Deviation | Range | Standard Deviation | Range | Standard Deviation | Range | Standard Deviation |
| Standard Calibration | [-0.2%;+0.1%] | 0.0003 | [-0.1%;+0.1%] | 0.0003 | [-0.3%;+0,2%] | 0.0004 | [-0.3%;+0.2%] | 0.0004 |
| Vega-Weighted Calibration | [-0.2%;+0.1%] | 0.0003 | [-0.1%;+0.1] | 0.0002 | [-0.1%;+0.1] | 0.0004 | [-0.1%;+0.1%] | 0.0004 |

**Table 6**: SABR model calibration errors over all the market volatility smile. For each calibration procedure (standard and vega-weighted) and for each valuation date (31st March and 31st August 2010), we report the range of minimum and maximum calibration errors and the standard deviation of the errors (equally-weighted for standard calibration and vega-weighted for vega-weighted calibration).

Overall, the SABR model performs very well at both dates with both pricing methodologies. In particular, we notice that in the short term (2-year, upper panels in Figure 10 and Figure 11) the standard SABR calibration (red line) seems, at first sight, closer to the market volatility (blue dots) and to better replicate the trend in the OTM regions. However, a closer look reveals that there are significant differences in the ATM area, where even small calibration errors can produce sensible price variations. Instead, the vega-weighted SABR calibration (green line) gives a better fit of the market volatility smile in the ATM region, in correspondence of the maximum vega sensitivity, and allows larger differences in the OTM regions where the vega sensitivity is close to zero. Thus the vega-weighted calibration permits a more efficient fit in the volatility surface regions that are critical for option pricing. The effects is less visible for long terms (middle and lower panels in Figure 10 and Figure 11) because of the higher vega sensitivity in the OTM regions. The closer replication of the vega-weighted SABR in the short term can be observed even in Figure 12 and in Figure 13.

Both the standard and the vega-weighted approaches lead to similar results in terms of range of minimum and maximum errors and standard deviation (see Table 6). In particular, the standard deviation measures of the errors over the 30-year term structure are almost the same: this is due to the fact that only in the short term (up to 4 years) the two calibration differ and using a vega-weighted minimization can ensure a more better fitting of the market data, as shown in the upper panels of Figure 10 and Figure 11.

We conclude that the SABR model is quite robust under generalisation to the modern pricing framework and can be applied to properly fit the new dynamics of the market volatility smile and to price off-the-market options coherently with the new market evidences.

## 5. Conclusions

In this work we have presented a quantitative study of the markets and models evolution across the credit crunch crisis. In particular, we have focused on the fixed income market and we have analyzed the most relevant empirical evidences regarding the divergences between Libor vs OIS rates, between FRA vs forward rates, the explosion of Basis Swaps spreads, and the diffusion of collateral agreements and CSA-discounting, in terms of credit and liquidity effects. These market frictions have induced a segmentation of the interest rate market into sub-areas, corresponding to instruments with risky underlying Libor rates distinct by tenors, and risk free overnight rates, and characterized, in principle, by different internal dynamics, liquidity and credit risk premia reflecting the different views and preferences of the market players.

In response to the crisis, the classical pricing framework, based on a single yield curve used to calculate forward rates and discount factors, has been abandoned, and a new modern pricing approach has prevailed among practitioners, taking into account the market segmentation as an empirical evidence and incorporating the new interest rate dynamics into a multiple curve framework. The latter has required a deep revision of classical no-arbitrage pricing formulas for plain vanilla interest rate derivatives, now funded on the risk neutral measure associated to the risk free bank account and on the martingale property of the FRA rate under such measure. In particular, we have reported the multiple-curve generalization of the SABR model, the simplest extension of the well known Black's model with stochastic volatility, routinely used by market practitioners to fit the interest rate volatility smile and to price vanilla Caps/Floors and Swaptions.

In section 4 we have reported the results of an empirical analysis on recent market data comparing three different pre- and post-credit crunch pricing methodologies and showing the transition of the market practice from the classical to the modern pricing framework. In particular, we have proven that the market of Interest Rate Swaps since March 2010 has abandoned the classical Single-Curve pricing methodology, typical of the pre-credit crunch interest rate world, and has adopted the modern Multiple-Curve CSA approach, thus incorporating into market prices the credit and liquidity effects. The same happened with European Caps/Floors, with the full transition to the CSA-discounting methodology retarded up to August 2010. Finally, we have proven that the SABR model is quite robust under generalisation to the modern pricing framework and can be applied to properly fit the new dynamics of the market volatility smile and to price off-the-market options coherently with the new market evidences.

The work presented here is a short step in the long-run theoretical reconstruction of the interest rate modelling framework in a post-crisis financial world, with Libor rates incorporating credit and liquidity risks. We believe that such risks and the corresponding market segmentation expressed by large basis swap spreads will not return negligible as in the pre-crisis world, and will be there in the future, exactly as the volatility smile has been there since the 1987 market crash. Expected future developments will regard, for example, the extension of pre-crisis pricing models to the Multiple-Curve world with stochastic basis, and the pricing of non-collateralized OTC derivatives including consistently the bilateral credit risk of the counterparties in the form of Credit Value Adjustment (CVA) and Debt Value Adjustment (DVA), and the liquidity risk of the lender in the form of Liquidity Value Adjustment (LVA) (see e.g. Bianchetti and Morini (2010)).